\documentclass[12pt,final,nofootinbib]{revtex4}
\usepackage{hyperref}
\usepackage{amsmath,amsfonts,amssymb,color,bbm,bm}
\usepackage{xspace,graphics,epsfig,multirow,soul}
\def\del{\partial}

\def\lb{\left(}
\def\rb{\right)}
\def\ls{\left[}
\def\rs{\right]}
\def\lc{\left\{}

\def\nn{\nonumber}
\def\rc{\right\}}

\def\baray{\begin{eqnarray}}
\def\earay{\end{eqnarray}}

\def\i{{\mathbbm i}}
\newcommand{\set}[1]{{\{#1\}}}
\newcommand{\be}{\begin{equation}}
\newcommand{\ee}{\end{equation}}
\newcommand{\bea}{\begin{eqnarray}}
\newcommand{\eea}{\end{eqnarray}}

\def\lsim{\mbox{\raisebox{-.6ex}{~$\stackrel{<}{\sim}$~}}}
\def\gsim{\mbox{\raisebox{-.6ex}{~$\stackrel{>}{\sim}$~}}}
\begin{document}
\title{Kaluza-Klein relics from warped reheating}

\author{Aaron Berndsen, James M.\ Cline}
\affiliation{%
\centerline{Physics Department, McGill University,
3600 University Street, Montr\'eal, Qu\'ebec, Canada H3A 2T8}
e-mail: aberndsen@hep.physics.mcgill.ca, 
jcline@physics.mcgill.ca }
\author{Horace Stoica}
\affiliation{
\centerline{Blackett Laboratory, Imperial College, London SW7 2AZ,
U.K.}e-mail: f.stoica@imperial.ac.uk}

\date{\today}

\begin{abstract} 
\thispagestyle{plain}

It has been suggested that after brane-antibrane inflation in a
Klebanov-Strassler (KS) warped throat,  metastable Kaluza-Klein (KK)
excitations can be formed due to nearly-conserved angular momenta
along isometric directions in the throat.  If sufficiently
long-lived, these relics could conflict with big bang
nucleosynthesis or baryogenesis by dominating the energy density of
the universe.    We make a detailed estimate of the decay rate of
such relics using the low energy effective action of type IIB string
theory compactified on the throat geometry, with attention to powers
of the warp factor.  We find that it is necessary to turn on
SUSY-breaking deformations of the KS background in order to ensure
that the most dangerous relics will decay fast enough.  The decay
rate is found to be much larger than the naive guess based on the
dimension of the operators which break the angular isometries of the
throat. For an inflationary warp factor of order $w\sim 10^{-4}$, we
obtain the bound $M_{3/2} \gsim 10^9$ GeV on the scale of SUSY
breaking to avoid cosmological problems from the relics, which is
satisfied in the KKLT construction assumed to stabilize the
compactification.  Given the requirement that the relics decay
before nucleosynthesis or baryogenesis, we place bounds on the mass
of the relic as a function of the warp factor in the throat for more
general warped backgrounds.   
\end{abstract}

\maketitle

\section{Introduction}
\label{sec:WR_intro}
The success of the inflationary paradigm in providing a natural
resolution for the flatness and homogeneity problems of Standard Big
Bang cosmology has made inflation an essential part of early-universe
cosmology. This success has led to intense efforts in realizing
inflation within string theory, resulting in several new scenarios
including brane-antibrane inflation  \cite{KKLMMT}, where the
interbrane separation plays the role of the inflaton. These
constructions provide new possibilities for constraining the
parameters of string theory, within compactifications that could be
compatible with the Standard Model.

A potential source of new phenomenological constraints, distinct from
those arising in generic field theory models of inflation, is the
reheating process at the end of inflation.  If inflation occurs in
one warped throat, while the standard model (SM) is localized in
another throat, there can be a difficulty in transferring the energy
from brane-antibrane annihilation to the standard model degrees of
freedom since the warp factor provides a gravitational potential
barrier between the throats.   If the barrier cannot be penetrated,
``reheating'' will be predominantly into invisible gravitons
\cite{KL}-\cite{BC}, an unacceptable outcome.  In ref.\
\cite{bib:barn} it  was argued that the suppression due to the
barrier can be counteracted by the enhanced coupling of the
Kaluza-Klein (KK) modes to the deeply--warped SM throat; this
scenario has been further studied in \cite{Chialva}-\cite{Lang}.
Another interesting possibility is that inflation could deform the SM
throat in such a way that its oscillations at the end of inflation
efficiently reheat the SM degrees of freedom \cite{Frey}.

A further challenge was recently pointed out in
ref.~\cite{bib:kofmanyi}, whose authors highlighted the possibility of
long-lived, heavy KK modes,  which could conflict with standard cosmology.  In
previous studies of reheating in a warped throat, the throat was modelled by a
single extra dimension, leading to an AdS$_5$ geometry. Massive states are
strongly peaked in the infrared (IR) region of the throat, so integrating them
out by  dimensional reduction (DR) resulted in large effective couplings, and
hence efficient decay. Ref.\  \cite{bib:kofmanyi} emphasized that the actual
background solution, the Klebanov-Strassler (KS) solution \cite{KS}, contains an
additional 5D internal space $\mathcal{M}_5$ with isometries along which
nonradial KK excitations can occur. For realistic particle phenomenology
$\mathcal{M}_5$ is usually taken as $T^{1,1}$, the Einstein-Sasaski manifold
for the group ${SU(2)\times SU(2)}/{U(1)}$. These isometries, as we shall
review, result in approximately conserved angular momenta which constrain the
possible decay channels and result in a long-lived relic corresponding to the
lightest\footnote{As we will discuss, there are also zero-mass charged
states in the KS background.  Here we refer to the lightest massive
KK states.} 
 ``charged'' state, {\it i.e.,} the lightest state with angular
momentum in the $T^{1,1}$. We shall refer to this candidate relic as the {\it
lightest {\it massive} charged state} (LMCS).

If the KS throat was the entire compactification manifold, the angular
isometries would be exact and the LMCS would be stable.  However it is necessary
to cut off the throat in the ultraviolet (UV) region, joining it to a larger
Calabi-Yau (CY) manifold which does not globally  preserve the isometries.  The
process of gluing together the KS throat to the CY thus perturbs the KS geometry
in the UV region, and this information propagates down the throat into the IR. 
We will assume that a mode of the metric which was zero due to the isometry will
be sourced in the UV region, and that its radial profile decays exponentially
toward the IR region so that the symmetry breaking is a weak effect in the IR. 
In the CFT description, via the AdS/CFT correspondence, this corresponds to
turning on an irrelevant operator that breaks the symmetry.  Of course, if the
operator is relevant the symmetry-breaking is strong in the IR, and there is no
problem of long-lived relics  since in this case the throat geometry is not
close to the KS solution. 

Since the symmetry-breaking effect is suppressed in the IR, while the
radial profile of the LMCS is strongly peaked in the IR, the
operators induced in the low-energy effective theory that describe
the decay of the LMCS will be  suppressed by powers of the warp
factor $w$, which determines the hierarchy of scales between the
bottom and top of the throat.  If this suppression were too strong,  
the heavy KK relics could be long-lived and come to dominate the
energy density of the universe at unacceptably high temperatures.  In
particular, they should decay before the era of big bang
nucleosynthesis at $T\sim 1$ MeV at the very least, and most likely
also before baryogenesis, since otherwise the entropy produced by
their decays will greatly dilute the baryon asymmetry. Assuming that
baryogenesis could not have happened later than the electroweak phase
transition requires the KK relics to decay at temperatures greater
than 100 GeV. We will show that to avoid this in the $T^{1,1}$
background, a SUSY breaking operator must be turned on at a scale
greater than $6\times 10^8$ GeV (more generally, $100\, w^{-1.7}$ GeV for
warp factor $w$).  In addition, even if SUSY is broken at  the
Planck scale (or if the KK relics can decay without breaking SUSY in
backgrounds other than KS), we derive general constraints which  can
be used to further exclude backgrounds with a very massive LMCS or a
large warp factor.

In ref.\  \cite{bib:kofmanyi}, it was assumed that the suppression of
the LMCS decay amplitude was of the form $w^p$, where $p+4$ was the
dimension of the most relevant charge-violating (but 5D Lorentz and
SUSY preserving) operator in the CFT.   However, there was no
detailed justification for this  assumption, and it is not obvious
that it should give the same answer as actually computing the decay
rate from the effective theory.  Our goal in this paper is to  make
an accurate estimate of the decay rate of the potentially dangerous
relics in the $AdS_5\times T^{1,1}$  type IIB supergravity
background.  We will find a parametrically different result 
than that of ref.\ \cite{bib:kofmanyi}.  Moreover, we will show that the CFT
operator considered in ref.~\cite{bib:kofmanyi} is not sufficient to
destabilize the LMCS in the KS background: one must turn on, in
addition, an irrelevant SUSY-breaking operator for this purpose. 
We also show that decays are possible without invoking SUSY-breaking,
but we did not find any example which gave a sufficiently large decay
rate.

 In ref.\ \cite{Chen:2006ni} it was found that the density of KK
relics could be suppressed to an acceptable level through
annihilations (rather than the decays we study in this paper)  if
the warp factor $w$ is sufficiently small, whereas larger values are
needed for sufficiently fast decays.  In section \ref{comparison} we
will compare the two approaches, noting that the values of $w$
needed for annihilations to sufficiently deplete the KK relics are
much smaller than one would like for getting the right
inflationary scale in the throat.

In the remainder of the paper we examine the constraints on the 
warp factor and the mass of the LMCS resulting from considerations
of BBN and baryogenesis, both for the KS background, and for more
general warped compactifications. Section \ref{sec:bkgnd} gives a
brief introduction to the problem at hand, including the origin of
the long-lived relic and the means through which it may decay. For
simplicity, section \ref{sec:bkgnd} employs a toy model that
captures the key ingredients of the analysis, while section
\ref{sec:kkdisc} analyzes the problem of the LMCS decay in the 10D
supergravity background. This includes identifying the LMCS (section
\ref{sec:lmcs}) and its interactions (section \ref{sec:ilmcs}), and
deriving constraints on the scale of SUSY breaking (section
\ref{sec:decaychannel}) as well as more general constraints in
warped models where the LMCS mass may take different values (section
\ref{sec:general}).  In section  \ref{sec:susydecay} we identify a
decay channel which does not require any SUSY-breaking background,
but we find that it is too suppressed to allow for sufficiently fast
decay of the LMCS.  In section  \ref{comparison} we compare the
decay scenario to that where the KK relics annihilate with each
other, showing that annihilations of the LMCS are too inefficient
unless the warp factor is $w\sim 10^{-8}$, which is far below the
value needed for brane inflation. We give conclusions in section
\ref{sec:discussions}.

Some details are reserved for the appendices, including a
comprehensive description of the radial behaviours for 5D scalars, 
1-forms and antisymmetric 2-forms (Appendix~\ref{app:app2}), and a
discussion of the $T^{1,1}$ harmonics (Appendix~\ref{sec:angints}).

\section{Background Deformations and KK Mode Decay}
\label{sec:bkgnd}
In this section we will describe in greater detail the origin of
the symmetry breaking for the approximate angular isometries, and
we will illustrate the approach we are going to take using a 
simplified toy model. 

The problem of relic angular KK modes is closely related to a moduli
problem associated with having an anti-D3 brane
$\left(\overline{D3}\right)$ placed at the bottom of the deformed 
conifold geometry, as one might wish to do in order to uplift the 
AdS vacuum from K\"ahler modulus stabilization to dS or Minkowski 
space, in the manner of KKLT \cite{KKLT}.   The energy of the 
$\overline{D3}$ is minimized at the bottom of the throat, but, as
pointed out in ref.\  \cite{bib:aharony} the base of the deformed
conifold has an $S^{2}\times S^{3}$  topology, whose $S^2$ shrinks to
vanishing size at the  location of the $\overline{D3}$. The
$\overline{D3}$ can move freely inside the $S^{3}$, whose coordinates
thus correspond to 3 massless moduli. In order to stabilize  these
moduli one needs to break the isometries of the $S^{3}$.  

In ref.\ \cite{bib:aharony} this was achieved by considering the dual
field theory to the KS background and turning on an irrelevant
operator that gives a mass to the fields describing the
$\overline{D3}$ position.  The same mechanism might also destabilize
the would-be angular KK relics since the operator provides a
background correction which perturbs the geometry and breaks the
symmetries. To analyze this process we choose to work in the gravity
side of the gauge/gravity correspondence. Since a renormalization
group (RG) flow in the field theory dual corresponds to movement
along the radial direction of the AdS space, turning on an operator
in the ultraviolet (UV) of the field theory and performing the RG
flow corresponds to turning on a source for the bulk classical field
dual to that operator and following its effect along the radial
direction  to the bottom of the throat. 

In the AdS background geometry the fields have an exponential
dependence on the radial direction, and the profile of the
symmetry-breaking perturbation will be related to the warping of the
background geometry. This perturbation is sourced by the  CY, which
generically does not preserve the symmetries of the throat; hence we 
consider a source in the UV which depends nontrivially on the 
angular coordinates of the $S^{3}$ cycle, and  thus breaks the
corresponding isometries of the $S^{3}$. As a consequence, the KK
modes of fields that couple to the source will become unstable since
the KK quantum numbers are no longer conserved quantities; 
this is just the gravitational dual  of the mechanism that made the
$\overline{D3}$ position moduli massive in  \cite{bib:aharony}.

\subsection{A simplified model}
\label{simp}

Our basic approach will be to compute the 4D effective Lagrangian
for KK relics in the presence of a perturbation to
the background geometry.  The perturbation leads to symmetry-breaking
terms in the effective Lagrangian, including vertices for the decay of
the relic.  It is useful to illustrate this procedure on a simpler
model before tackling the full 10D supergravity theory.

We therefore consider a massless scalar field $\phi$ in 6D, where one
of the compact dimensions corresponds to the angular direction which 
is an isometry of the unperturbed throat (in this case the $U(1)$ symmetry
along a circle of length $L$) and the other is the radial
direction along the AdS. Its Lagrangian is
\be
	{\cal L} = \frac12 \int d^{\,4}x\, dr\, d\theta \sqrt{-g}\,
	g^{AB}\partial_A\phi\, \partial_B\phi
\label{Ltoy}
\ee
and the original throat geometry is described by the line element
\be
	ds^2 = a^2(r)\,\eta_{\mu\nu} dx^\mu dx^\nu + dr^2 + L^2 d\theta^2
\label{dstoy}
\ee
between $r=0$ and $r=r_0$. 
In Randall-Sundrum coordinates \cite{RS}, the
warp factor takes the form $a=e^{-kr}$, where $k$ is the AdS scale; so
$r=0$ corresponds to the 
top of the throat (the UV), where it joins to the CY, and $r=r_0$ is the bottom
of the throat (the IR).  
To be a solution to Einstein's equations, 
this geometry requires an exotic bulk
stress energy, with $T^a_b = k^2$diag$(6, 6, 6, 6, 6, 10)$, but we
only use it as an illustrative toy model.
To model the symmetry-breaking effect of the CY, there are two
possible effects on the background.  One is that the metric
(\ref{dstoy}) gets perturbed by 
\be
	\Delta ds^2 = \sum_n\sin(n\theta)\left[
	a^2(r)\alpha_n(r) dx^2 + \beta_n(r) dr^2 + L^2 \gamma_n(r)d\theta^2
	\right]
\label{ddstoy}
\ee
corresponding to KK excitations of the angular direction.  (For
simplicity we ignore the fluctuations proportional to
$\cos(n\theta))$.)
Another possibility is that angular KK modes of the scalar $\phi$
get sourced,
\be
	\Delta\phi = \sum_n\sin(n\theta)\varphi_n(r)
\label{dphi}
\ee 
All these can be thought of as solutions to the vacuum 
Einstein or scalar field equations
in the throat, sourced by some boundary conditions at the CY, $r=0$.
In the AdS background, the solutions for the radial wave functions
generically have the form
\be
	\alpha_n(r) = \sum_{\pm} \alpha_{n\pm} e^{-z_{n,\pm} kr}
\label{fluct}
\ee
(similarly for $\beta_n$, $\gamma_n$, $\varphi_n$)
where $z_{n,+}$ and $z_{n,-}$ are related by
\be
	\Delta_n = -z_{n,-},\quad 4-\Delta_n = -z_{n,+}\,.
\ee
In the AdS/CFT, $\Delta_n$ is the dimension of an operator
${\cal O}_n$ which corresponds to the deformation (\ref{fluct}) of
the background geometry.

An important concept throughout this paper is the necessity for
perturbations like (\ref{fluct}) to remain small as one goes to the
bottom of the throat; otherwise the KS throat is not a good
approximation to the actual geometry.  This requirement is fulfilled
as long as we only turn on marginal or irrelevant operators in the 
CFT, with $\Delta_n \ge 4$.  In that case, $z_{n,+} < 0$ and the
corresponding solution decays toward the bottom of the throat. If
the operator is relevant, then both solutions grow toward the  IR,
and it is impossible to keep them small without fine-tuning in the
UV.  

As explained in ref.\ \cite{bib:aharony}, the coefficient of the
$z_{n,-}$ solution, which grows toward the IR, is not exactly zero,
but depends on the boundary conditions at the bottom of the throat,
$r=r_0$.  Regardless of the details of these boundary conditions
however, generically one expects both solutions to be of the same
order of magnitude at $r_0$.  As one moves toward the top of
the throat, the $z_{n,+}$ solution quickly comes to dominate.  For the
purposes of the kind of estimates we will be making, it is sufficient to
approximate the background deformations by just the $z_{n,+}$ part of
the solution.  Thus, for example, 
\be
	\alpha_n(r) \cong \alpha_n e^{(4-\Delta_n)kr} = \alpha_n
	e^{-z_{n,+}kr}\,,
\ee
where $\alpha_n$ characterizes the magnitude of the symmetry breaking
in the UV, and thus could be $\mathcal{O}(1)$.  	

Now we turn our attention to the propagating fluctuations of the
scalar field, which include the would-be stable relic KK excitation.
The  radial ($n$) and angular ($m$) KK decomposition is 
\be
	\phi = {1\over\sqrt{2\pi L}}\sum_{n,m} R_{nm}(r) e^{im\theta} 
	\phi_{nm}(x^\mu)\,.
\ee
In the absence of the metric and scalar
perturbations (\ref{ddstoy},\ref{dphi}), the interactions of the angular excited states with $n=0$, 
$m\neq 0$ conserve the total angular momentum, so there is no way 
for the massive states $\phi_{0,\pm 1}$ to decay.  These, then,
represent the lightest massive charged states (LMCS) in the toy model.

In the perturbed metric, angular momentum is no longer conserved, and
we can construct an interaction from the kinetic term for the decay
$\phi_{0\pm 1} \to \phi_{00} h_{\mu\nu}$, where $h_{\mu\nu}$ is a massless graviton
which is a perturbation  about the Minkowski metric factor
in (\ref{dstoy}): $\eta_{\mu\nu}\to \eta_{\mu\nu} + h_{\mu\nu}/M_p$.
This decay channel may come from the 6D kinetic term, and the 4D
effective interaction is 
\be
	{\cal L}_{\rm decay\ 1} = h^{\mu\nu}\,\partial_\mu\phi_{0\pm1}
	\,\partial_\nu\phi_{00}\,
	\, \left[-{\alpha_1\over 4\pi M_p}\int d\theta \sin\theta e^{\pm i\theta}
	\int dr\,  e^{-(2+z_{1,+})kr}\, R_{01}(r) R_{00}(r) \right]
\label{effint}
\ee
However, this particular decay channel is not illustrative of the more
realistic SUGRA theory we are ultimately interested in, because it
involves the massless scalar $\phi_{00}$ as a decay product.  In a
realistic theory, massless scalars should be somehow projected out
or given a mass, to avoid various phenomenological difficulties.

Another possibility, which does not require any massless
scalar mode, is to use the
scalar perturbation (\ref{dphi}) to the background.  Substituting
such a deformation for one of the $\phi$ factors in the kinetic term
(\ref{Ltoy}) allows us to generate a vertex for the $\phi_{0\pm1}$
excitation to decay into massless gravitons, $\phi_{0\pm1}\to h h$.
\bea
	{\cal L}_{\rm decay\ 2} = \phi_{0\pm1}
	h^{\mu\nu}h_{\mu\nu} {\varphi_1\over 8 M_p^2}\sqrt{L\over
2\pi} \!\!\!&
\Big[  \int d\theta \sin\theta e^{\pm i\theta}
	\int dr\,  e^{-4kr}\partial_r e^{-z_{1,+}kr} 
	\partial_r R_{01}(r) R^2_{00}(r)
\nonumber\\
 &\pm i L^{-2}\int d\theta \cos\theta e^{\pm i\theta}
	\int dr\,  e^{-4kr-z_{1,+}kr} 
	\,R_{01}(r) R^2_{00}(r)
\Big]
\label{effint22}
\eea
where we note that $R_{00}(r)$ is just a constant.  This process is
more analogous to the ones we will be interested in for the SUGRA
model, so we will focus on it.

To evaluate the 
radial integrals in (\ref{effint22}) we must solve for the radial wave functions, which 
obey the equation of motion  with 4D mass $m_{mn}$
\be
 e^{2kr}\del_{r}\lb e^{-4kr}\del_{r}R_{nm}\rb
- \frac{m^2}{L^2} e^{-2kr} R_{nm} + m_{nm}^2 R_{nm} = 0\,.
\label{eq:Rgen}
\ee
(see eq.~(\ref{a5}) for the generalization to 10D).
Defining the warp factor at the bottom of the throat
\be
	w = e^{-kr_0}\,,
\ee
the solutions have the form
\bea
R_{nm}(r)&\simeq&\frac{w\sqrt{k}e^{2kr}}{J_{\nu_m}(x_{nm})}\ls\,
J_{\nu_m}\lb x_{nm} we^{kr}\rb+w^{2\nu_m} Y_{\nu_m}\lb 
x_{nm}\,we^{kr}\rb\,\rs
\nn\\
&&\quad{\nu_m}=\sqrt{4+(m/kL)^2}\,,
\label{eq:RS}
\eea
where $x_{nm}\sim 1$ is determined by the boundary conditions, and the
4D mass of the excitation is given by $m_{nm} = k w x_{nm}$, 
as shown in ref.\ \cite{bib:goldwise}. 
 The radial behaviour for excited modes is dominated by
the $J_\nu$ solution, which is strongly peaked in the IR.  The
zero-mode solution $R_{00}\cong\sqrt{2k}$ is just a constant, whose
value is determined by the normalization
condition
\be
	\int dr e^{-2kr} R_{nm}^2 = 1\,.
\ee
The reader is referred to Appendix~\ref{app:app2} for a detailed
discussion of the radial behaviour. There, a discussion of vector
fields and antisymmetric tensor fields is also 
included since they have a different radial behaviour that
is important for the SUGRA theory we will discuss. 

We can estimate the integrals
determining the coefficient of the decay-mediating operator in
(\ref{effint22}) using the small- and large-$r$ asymptotics of the
Bessel function.  Near $r=0$, the argument of $J_\nu$ is
exponentially small and $J_\nu(x)\sim  (x/2)^\nu/\Gamma(1+\nu)$,
while near $r=r_0$, the argument is of order unity.  Both behaviours
are consistently approximated by $R_{01} \sim w^{1+\nu}\sqrt{k}
e^{(2+\nu)kr}$, leading to the estimate
\be {\cal L}_{\rm decay\ 2}
\cong 
\phi_{0\pm1}
	h^{\mu\nu}h_{\mu\nu}\, w^{1+\nu} {\varphi_1\over
8 M_p^2}\sqrt{Lk\over
2\pi} \left[\pm i(2+\nu)z_{1,+}k^2 - L^{-2}\right]
	\int dr e^{(\nu-2 - z_{1,+})kr}\,.
\label{effint2} 
\ee 

Our principal interest in the present work will be to track the
parametric dependence of the decay amplitude on the warp factor $w$,
since it can be the strongest source of suppression.  To complete the
evaluation of (\ref{effint2}), it is necessary to know whether the
combination $(\nu-2 - z_{1,+})$ appearing in the exponent is positive
or negative.  In the former case, the radial integral grows toward
the IR, and yields an inverse power of the warp factor,  $k^{-1}
w^{-(\nu-2 - z_{1,+})}$.  In the latter case, the integral converges
in the IR, and yields only the mild dependence 
$k^{-1}|\nu-2 - z_{1,+}|^{-1}$.  In the present example, we have not
explicitly determined the value of $z_{1,+}$ to determine which case
is true, but in the SUGRA model we will do so.  

Let us contrast this with the estimate made in ref.\ 
\cite{bib:kofmanyi}.  There it was assumed that the operator
${\cal O}_+$ (corresponding to a background correction to the metric)
directly mediates the decay in the 4D effective theory, and its degree
of irrelevance controlled the amount of warp factor suppression, so that 
\be
\label{guess}
	{\cal L}_{\rm decay} \sim w^{\Delta_+-4} = w^{z_{1,+}}\,.
\ee
On the other hand our explicit calculation indicates that the
warp factor dependence in ${\cal L}_{\rm decay}$ is either
$w^{1+\nu}$, or else $w^{3+z_{1,+}}$, depending on the sign of
$(\nu-2 - z_{1,+})$.

In the full 10D SUGRA model we want to consider, the simple decay
process illustrated above will not be possible because the background
deformation corresponding to $\varphi_1$ will turn out to be a relevant
operator in the CFT. This contradicts our requirement that only
deformations of the throat which decay toward the IR are allowed.
We will be able to overcome this by turning on a different irrelevant
deformation, which leads to mixing between $\phi_{0,\pm 1}$ and some
massless field which carries trivial angular quantum numbers.  The
heavy $\phi_{0,\pm 1}$ will thus appear as an intermediate state
in the decay process.
But the method illustrated in the previous subsection will still 
apply for computing the mixing amplitude.

Following the previous discussion, the radial
integral $\mathcal{R}$ appearing in the dimensional reduction from
10D to 4D of an interaction involving $N_i$ appearances of the
inverse 4D metric, $N_m$  massive modes,  $N_0$
massless modes, and $N_t$ tadpole insertions  in the background, with
dimensions $\Delta_b>4$ is
\bea
\mathcal{R}&\simeq&\int dr e^{-(4-2N_i)kr}
\ls\Pi_{i=1}^{N_m}\,R_{\nu_{i}}(r)\rs
R_0(r)^{N_0}\ls\Pi_{b=1}^{N_t}R_{t_b}(r)\rs\nn\\ &\simeq& w^{N_m}\int
dr e^{-(4-2N_i)kr}\, \ls\Pi_{i=1}^{N_m} e^{2kr}J_{\nu_i}\lb
x_nwe^{kr}\rb\rs\,\ls\Pi_{b=1}^{N_t}
e^{(4-\Delta_b)kr}\rs\nn\\
&\sim& w^{N_m+\sum_i\nu_i}
\int dr\, e^{(-4 + 2N_i + N + \sum_i\nu_i
-\sum_b(4-\Delta_b))kr}\nn\\
 &\sim&
\underbrace{
w^{N_m + \sum_i\nu_{i}}}_{{\rm UV-contrib}}
+\underbrace{w^{4-2N_i + N_m - N + \sum_b(4-
\Delta_b)}}_{\rm IR-contrib}
\,. \label{eq:dimred}
\eea
where we have defined $N = 2N_s + N_v$
in terms of the number $N_s$ ($N_v$) of scalars (vectors)
amongst the $N_m$ massive states.  Thus $N_m-N = N_s-N_a$,
where $N_a$ is the number of antisymmetric 2-form states within
$N_m$.  The relation between the values of $\nu_i$ for the external 
states and the 5D mass eigenvalues is given in appendix 
\ref{app:app2} and, generally, depends on the helicity of the state.  
The final estimate in (\ref{eq:dimred}) is found by determining which 
contribution dominates, the UV or IR; the correct answer is whichever
is largest.

\section{KK decay in Type IIB Supergravity}
\label{sec:kkdisc}
We now want to apply the methods illustrated in the 6D model to the
KS geometry sourced by a stack of D3-branes
in the 10D theory.  The line element is
\be
ds^2=H^{-1/2}\eta_{\mu\nu}dx^\mu dx^\nu+H^{1/2}\lb
dR^2+R^2ds^2_{T^{1,1}}\rb\,,
\ee
where
\be
H(R)=\frac{27\pi}{4R^4}{\alpha^\prime}^2g_sM\left[
K+g_sM\lb
\frac{3}{8\pi}+\frac{3}{2\pi}\ln\left(\frac{R}{R_{max}}\right)
\rb\right] 
\ee
is in terms of the flux quantum numbers $K$ and $M$.
Far from the tip $R<R_{max}$ we neglect the logarithmic
contributions to $H(R)$ and further ignore the small contribution
from the second term on the right to obtain the metric of an
$AdS_5\times T^{1,1}$ throat. Using the coordinate transformation
$R=k^{-1}e^{-kr}$, the metric becomes
\bea
ds^2&=&e^{-2kr}\eta_{\mu\nu}dx^\mu dx^\nu +
dr^2+\frac{1}{k^2}ds^2_{T^{1,1}}
\label{eq:ads}
\eea
The corresponding low-energy effective theory
is type IIB supergravity on an approximate
$AdS_5\times T^{1,1}$  background \cite{bib:johnson}:
\bea
S_{IIB}&=&\frac{1}{2\kappa_0^2}\int d^{10}x\sqrt{-G}
\lc e^{-2\phi}\ls R+4(\nabla\phi)^2-\frac{1}{12}\lb H^{(3)}\rb^2\rs
\right.\nn\\
&&\left.
-\frac{1}{12}\lb F^{(3)}+A^{(0)}\wedge H^{(3)}\rb^2-\frac12\lb
dA^{(0)}\rb^2
-\frac{1}{480}\lb F^{(5)}\rb^2
\rc
\nn\\
&&+\frac{1}{4\kappa_0^2}\int\lb A^{(4)}+\frac12B^{(2)}\wedge A^{(2)}\rb
\wedge F^{(3)}\wedge H^{(3)}\,.
\label{eq:iib}
\eea
As usual, $H^{(3)}$ is the field strength of the NS 2-form $B^{(2)}$,
$F^{(n+1)}$ is the field strength for the RR $n$-form $A^{(n)}$,
$2\kappa_0^2=(2\pi)^7{\alpha^\prime}^4$,
$\sqrt{\alpha^\prime}=l_s=M_s^{-1}$, where $M_s$ is the string scale.
In our subsequent analysis, we will refer to several other mass
scales.  The AdS curvature scale $k$  is determined by the flux 
quantum numbers $M$ and $K$ through the relation 
$k^{-4}\equiv\frac{27\pi}{4}{\alpha^\prime}^2g_sMK$.  The 
warped string scale, $w M_s$, is also determined by the fluxes,
through $w = e^{-2\pi K/(3g_s M)}$.  Finally, the Planck scale is
given by
$M_{p}^2=\frac{2V_6}{g_s^2\kappa_0}$ where $V_6$ is the
compactification volume.
From eq.\ (\ref{eq:ads}) it is apparent $V_6\simeq R_{AdS}^{6}=k^{-6}$,
so with $g_s<1$ and $k<M_s$ we find the 4D Planck scale is greater
than the string scale.

\subsection{Identifying the LMCS}
\label{sec:lmcs}
The first step is to discover which angular KK excitation, 
among the many
fields that result from dimensionally reducing the action
(\ref{eq:iib}), is the potentially dangerous relic, the LMCS.
Fortunately, the masses and quantum numbers of all the lowest-lying
KK excitations in the KS background have been tabulated in refs.\ 
\cite{bib:aharony,bib:ceresole}.  The correspondence between 10D
and 5D fields from integrating over the $T^{1,1}$ directions is
indicated in Table~\ref{tab:harmonics}, taken
from~\cite{bib:ceresole}.

\begin{table}[!tbh]\centering
\begin{tabular}{|c|c|c|c|c|c|c|}\hline
Dim&\multicolumn{5}{|c|}{fields}&harmonic\\
\hline\hline
10D&$h_{\mu\nu}(x,y)$&$h^a_a(x,y)$&$A_{abcd}(x,y)$&$B(x,y)$&$A_{\mu\nu}(x,y)$&\\ 
5D&$H_{\mu\nu}(x)$&$\pi(x)$&$b(x)$&$B(x)$&$a_{\mu\nu}(x)$&$Y(y)$\\
\hline
10D&$h_{a\mu}(x,y)$&$A_{\mu abc}(x,y)$&$A_{\mu a}(x,y)$&&&\\
5D&$B_{\mu}(x)$&$\phi_\mu(x)$&$a_\mu(x)$&&&$Y_a(y)$\\
\hline
10D&$A_{\mu\nu ab}(x,y)$&$A_{ab}(x,y)$&&&&\\
5D&$b^{\pm}_{\mu\nu}$&$a(x)$&&&&\\
\hline
10D&$h_{ab}(x,y)$&&&&&\\
5D&$\phi(x)$&&&&&$Y_{(ab)}(x,y)$\\
\hline\hline
10D&$\lambda(x,y)$&$\psi_{(a)}(x,y)$&$\psi_\mu(x,y)$&&&\\
5D&$\lambda(x)$&$\psi^{(L)}(x)$&$\psi_\mu(x)$&&&$\Xi(y)$\\
\hline
10D&$\psi_a(x,y)$&&&&&\\
5D&$\psi^{(T)}(x)$&&&&&$\Xi_a(y)$\\
\hline
\hline
\end{tabular}
\caption[The harmonic expansion of the 10D fields]{The harmonic expansion of the 10D fields. $h_{MN}$ is the 10D
metric, $A_{MNOP}$ the 10D four-form, $B$ the complex 0-form, and
$A_{MN}$ the 10D complex 2-form. We have not included the NS
2-form. The different polarizations of the fields appear as 5D scalars,
vectors, and tensors. (Adapted from ref.~\cite{bib:ceresole})}
\label{tab:harmonics}
\end{table}

The corresponding expansions of the fields in terms of scalar 
($Y^\set{\nu}$),
vector ($Y_a^\set{\nu}$) and tensor 
($Y_{ab}^\set{\nu}$) harmonics of the $T^{1,1}$ are given by
expressions like
\bea
\label{hexp}
h_{\mu\nu}(x,y)&=&\sum_\set{\nu}H_{\mu\nu}^\set{\nu}(x) Y^\set{\nu}(y)\\
\label{Bhexp}
h_{\mu a}(x,y)&=&\sum_\set{\nu}B^\set{\nu}_\mu (x) Y_a^\set{\nu}(y)\\
\label{habexp}
h_{(ab)}(x,y)&=&\sum_\set{\nu}\phi^\set{\nu}(x)Y^\set{\nu}_{(ab)}(y)\\
h^a_a(x,y)&=&\sum_\set{\nu}\pi^\set{\nu}(x)Y^\set{\nu}(y)\\
A_{abcd}(x,y)&=&\sum_\set{\nu}b^\set{\nu}{\epsilon_{abcd}}^e\mathcal{D}_eY^\set{\nu}\\ 
\label{bexp}
A_{\mu bcd}(x,y)&=&\sum_\set{\nu}\phi_\mu^\set{\nu}{\epsilon_{bcd}}^{ef}
\mathcal{D}_eY_f^\set{\nu}\,\label{lastharm}
\label{eq:harmonics}
\eea
where $x = (x^\mu,r)$, and $\set{\nu} = (j,l,r)$ are the quantum numbers
identifying the $T^{1,1} = SU(2)\times SU(2)/U(1)$ representation.  $j$
and $l$ are the usual angular momentum quantum numbers corresponding to
the two $SU(2)$ factors. Higher-rank fields are given in terms of their
dual representation. In general, $r = j_3-l_3$,
where $j_3,l_3$ are the respective
eigenvalues of the $T_3$ generators of the first and second
$SU(2)$'s.
For the scalar harmonics which will be of most
interest to us, $r = 2j_3=-2l_3$ ,
and so is restricted to the range $|r| < {\rm min}(2j,2l)$.

\begin{figure}[!bth]
\centerline{\includegraphics[width=0.5\textwidth]{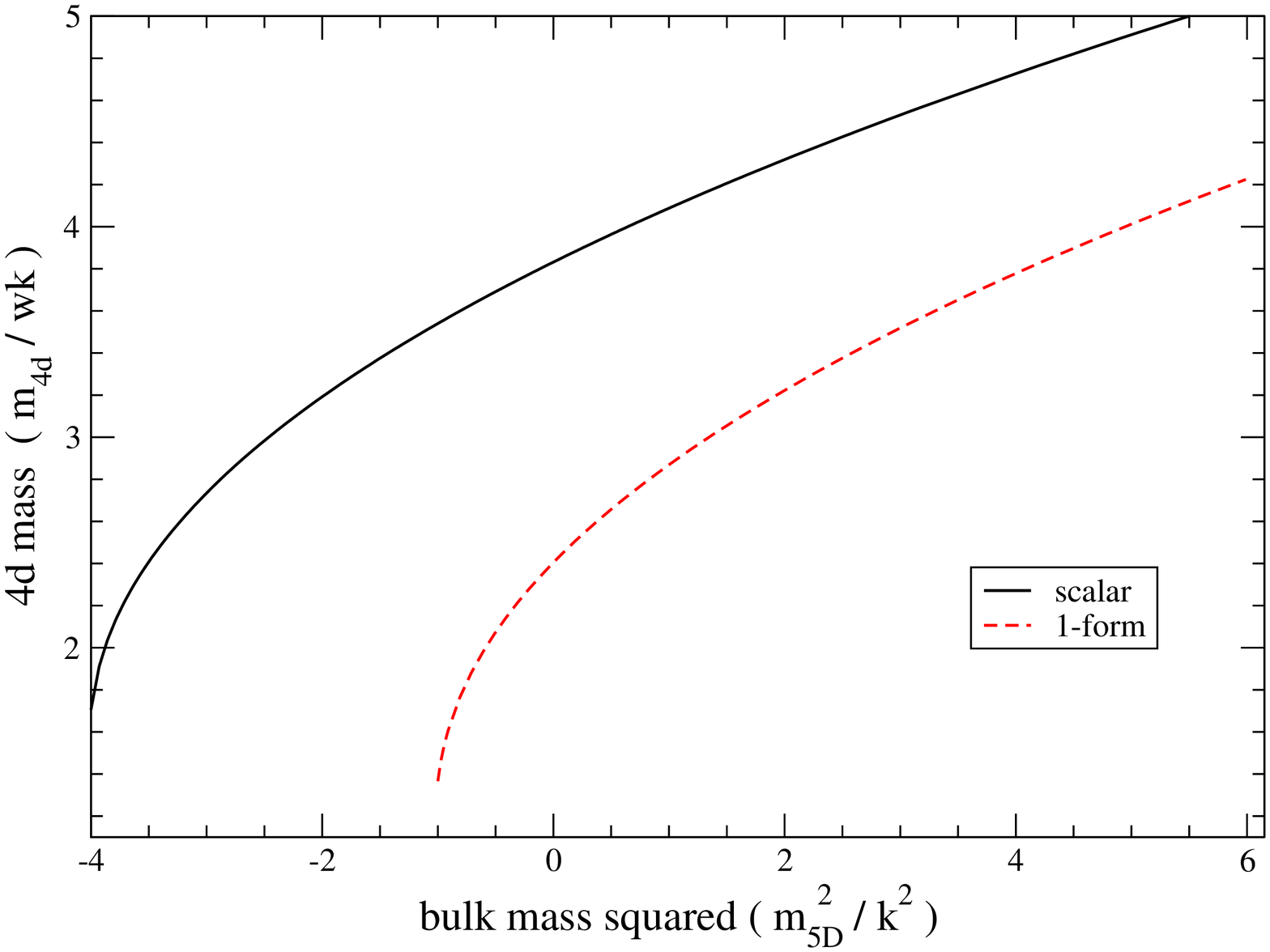}
\includegraphics[width=0.5\textwidth]{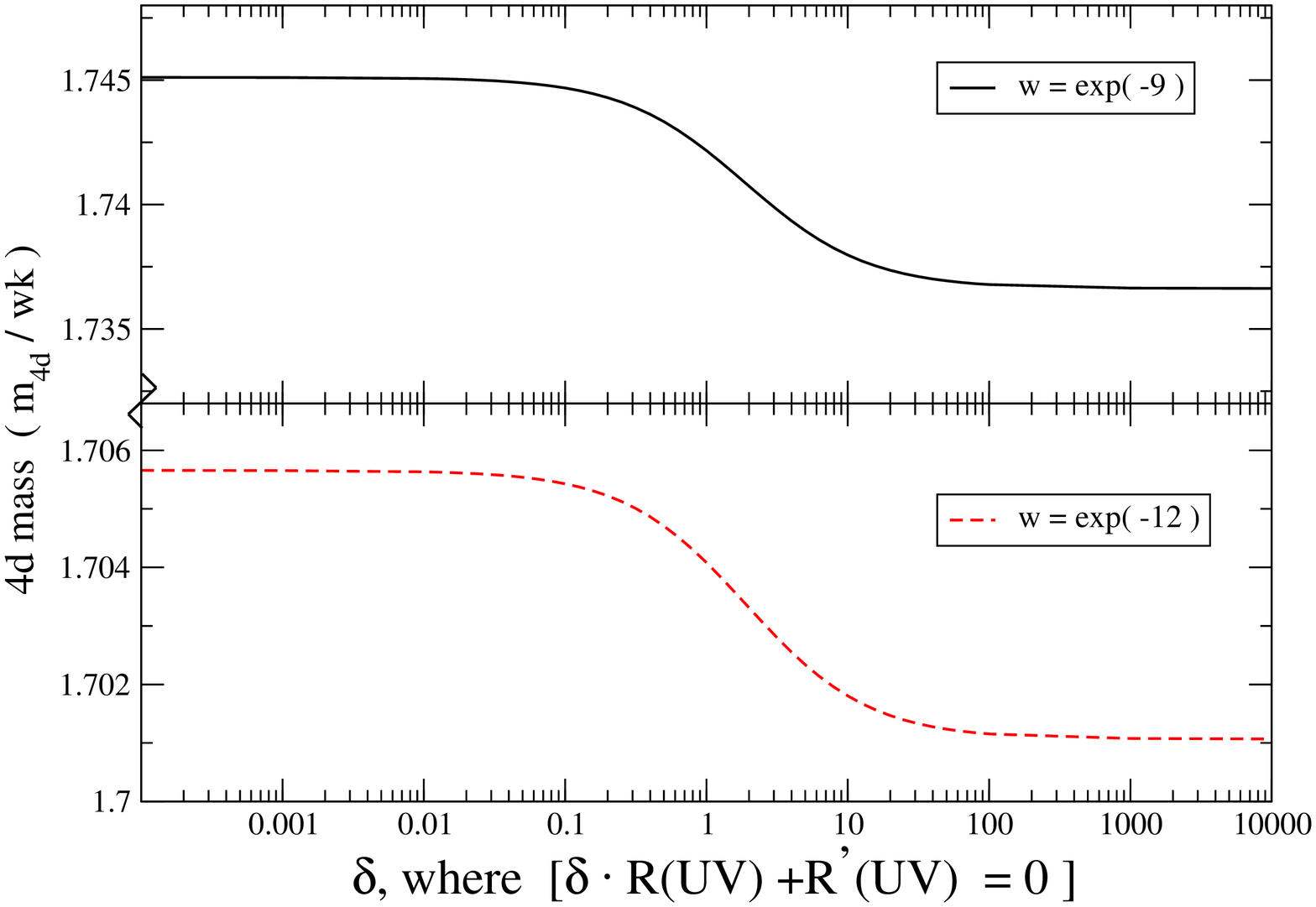}}
\caption{(a) Left: 4D mass of the first KK state as a function of the  5D
mass (squared). The result is presented for both 5d scalars and
vectors. (b) Right: dependence of the 4D mass on the UV boundary 
conditions, for two different values of the warp factor.}
\label{fig:m5vsm4}
\end{figure}

Ref.\ \cite{bib:ceresole} has computed the 5D masses of all the
states in the theory as functions of their $(j,l,r)$ quantum
numbers, and organized them into supermultiplets of the $N=1$
5D SUGRA theory. 
By going through these results and computing the masses of all
particles which have nontrivial $(j,l,r)$ values, we find that  the
field $b(x)$ in vector multiplet I is the LMCS,  with 
$(j,l,r)\in\{(1,0,0),(0,1,0)\}$. Thus there are two species of
LMCS, depending on whether $j$ or $l$ is nonzero.
 The 5D masses are defined in terms
of the ubiquitously appearing function
\be
	H_0(j,l,r) = 6\left( j(j+1)+l(l+1) - {r^2\over 8}\right)
\ee
which takes the value $H_0(1,0,0) = 12$ for the LMCS.  Its 5D mass is
given by
\be
	m^2_b = H_0 + 16 - 8\sqrt{H_0+4} = -4
\label{m2b}
\ee
in units of the AdS curvature, $k^2$.  As is well known 
\cite{witten}, squared
masses can be negative on an AdS background without leading
to instabilities, down to the 
 Breitenlohner-Freedman bound $m^2\ge-4$
\cite{BF}.  This bound is saturated by (\ref{m2b}).  To find the corresponding mass in 
4D, one must do the final dimensional reduction on $r$ by solving for
the radial wave functions. These are identical to the 6D case
(\ref{eq:RS}) (assuming that we approximate the throat
geometry by AdS$_5$), except for the replacement of the 5D mass,
$m_b^2$, in the index of the Bessel function
\be
	\nu = \sqrt{4 + m^2_b}\,.
\ee
We have carried out this calculation to find the 4D mass as a
function of the 5D one, using the boundary conditions of the RS
model, as in ref.\ \cite{bib:goldwise}.  The result, shown in Fig.\ 
\ref{fig:m5vsm4}(a) shows that the 4D LMCS mass is approximately
$m_{4D} = 1.7 w k$. Fig.\ \ref{fig:m5vsm4}(b) shows this value is 
quite insensitive to the details of the boundary conditions in the UV
(whether they are Neumann, Dirichlet, or mixed), which represent how
the throat is joined to the CY. Additionally, Fig~\ref{fig:m5vsm4}
extends the result found in ref.~\cite{bib:goldwise} from
$m_{5D}^2\ge0$ to include the Breitenlohner-Freedman range
$m_{5D}^2\ge-4$. It indicates the LMCS in 5D is  the LMCS in 4D as
well, {\it i.e.}, $m^2_{4D}$ is a monotonically increasing function
of $m^2_{5D}$. Notice that these results are for the lowest state in
the radial KK tower.  

\begin{figure}[!bth]
\centerline{\includegraphics[width=0.65\textwidth]{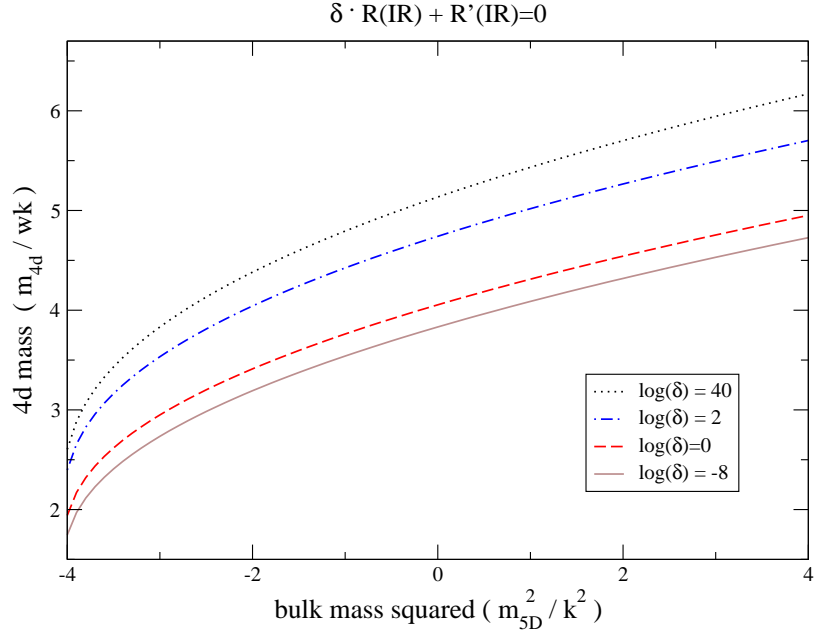}}
\caption{4D mass of the first KK state as a function of the  5D
mass (squared), for a wide range of b.c.'s of the form $R(r_0)\delta +
R'(r_0) = 0$ at the IR end of the throat, showing that $m_{4D}$ is
monotonically related to $m^2_{5D}$ regardless of the value of 
$\delta$.}
\label{bcfig}
\end{figure}

Of course the KS geometry is not exactly the same as the
Randall-Sundrum model, and the boundary conditions (b.c.'s) which should be
imposed at the bottom of the throat may not coincide exactly with the
$Z_2$ orbifold b.c.'s imposed in the RS model.  However,
this detail is not important for our estimates, and for the proper
identification of the LMCS, we need only rely on the fact that
regardless of the exact b.c.'s in the IR region, the relation between
the 4D mass and the 5D $m^2$ eigenvalue is a monotonic one.  This
insures that the particle with the lowest $m^2_{5D}$ will also be the
lightest one in the 4D effective theory.  We have verified the
monotonic property by for the entire range of mixed b.c.'s which
interpolate between Dirichlet and Neumann, as illustrated in fig.\
\ref{bcfig}.  

Table~\ref{tab:LMCS} lists the lowest-mass states in the spectrum.
Several saturate the Breitenlohner-Freedman bound but most of these
are uncharged; additionally, we identify the field $b(x)$ from the
4-form as the lightest charged state. Interestingly, most of the
lightest charged states come from the $4$-form, or the 10D graviton.
This table is useful for identification of the LMCS and
possible irrelevant operators that may be turned on in the background
to accommodate the decay.

\begin{table}[!tb]\centering%
\begin{minipage}{.49\textwidth}
{\small States corresponding to relevant operators:}
\begin{tabular}{|c|c|c|c|c|}\hline
Field&Multiplet&(5d) Mass&$\Delta$&QN's\\\hline
$b$&VM I&-4&2&(0,1,0), (1,0,0)\\
$\phi_\mu$&VM I&0&3&(1,0,0), (0,1,0)\\
$b$&VM I&-3&3&(1,1,$\pm 2$)\\
$\phi_1$, $\phi_2$&VM I&-3&3& (1,0,0), (0,1,0)\\
$b$&VM I&-2.33&3.29&(1,1,0)\\
\hline
\hline
\end{tabular}
\end{minipage}
\begin{minipage}{.49\textwidth}
{\small States of marginal and irrelevant
operators:}\\
\begin{tabular}{|c|c|c|c|c|}\hline
Field&Multiplet&(5d) Mass&$\Delta$&QN's\\\hline

$\phi_1/\phi_2$&VM I&0&4&(1,1,$\pm2$)\\
$\phi_3$&VM I&0&4&(1,0,0), (0,1,0)\\
$\phi_1/\phi_2$&VM I&1.25&4.29&(1,1,0)\\
$b$&VM I&1.40&4.32&(2,0,0), (0,2,0)\\
$a_1$&VM III/IV&2.79&4.61&(1,0,0), (0,1,0)\\
\hline
\end{tabular}
\end{minipage}
\caption[A list of the lightest states in the KS
 background.]{The lightest states charged in the KS background:  we list the 5D field,
 its supermultiplet, bulk mass, conformal dimension, and $T^{1,1}$
 quantum numbers. Most of the light, charged states correspond
 to the 4-form polarized along the $T^{1,1}$, $A_{abcd}^{(4)}$. The
 reader is referred to the appendices of~\cite{bib:ceresole} for the
 multiplet and mass listings.}
\label{tab:LMCS}
\end{table}

\subsection{Interactions of the LMCS}
\label{sec:ilmcs}
We have seen that in terms of the 10D fields,  the LMCS is contained
in the RR 4-form $A^{(4)}_{abcd}$ polarized along the internal
$T^{1,1}$ directions, resulting in  a massive scalar field
$b(x^\mu,r)$ from the 5D point of view.\footnote{Our conventions for
indices are: capital Latin letters $\{M,N,\ldots\}$ run over all
directions, small Latin letters $\{a,b,\ldots\}$ run over internal
directions, small Greek letters $\{\mu,\nu,\ldots\}$ run over the
four noncompact dimensions}  Therefore the relevant terms in the
action for type IIB supergravity~(\ref{eq:iib}) which provide 
interactions for the 4-form are
\bea
\label{action}
S_{IIB}(A^{(4)})&=&\frac{1}{2\kappa_0^2}\int
  d^{10}x\sqrt{-G}\ls-\frac{1}{240}\lb
  F^{(5)}\rb^2\rs+\frac{1}{2\kappa_0^2}\int A^{(4)}\wedge
  F^{(3)}\wedge H^{(3)}\,.\nn\\\label{eq:4formints}
\eea

Let us now try to follow the example of subsection \ref{simp} 
by turning on the CFT operator
used by ref.\ \cite{bib:aharony} to stabilize the $\overline{D3}$
moduli.  As shown there, this operator corresponds to a KK mode of the
warped metric on the $T^{1,1}$ with quantum numbers $(1,1,0)$, call it 
$\delta g_{ab}^{(1,1,0)}$.   In combining this operator with the LMCS
in the kinetic term for $A_{abcd}$, the way to make a $T^{1,1}$ 
singlet combination which is closest to the 6D example is to
choose the two different species of LMCS for the $A_{abcd}$ factors:
\be
	{\cal L}_{\rm int} =
	\int dr\, d^{\,5} y \sqrt{g_6}\,
	 \delta g^{aa'}_{(1,1,0)}\, h^{\mu\nu}\,
	\partial_\mu A_{abcd}^{(0,1,0)}
	\partial_\nu A_{a'b'c'd'}^{(1,0,0)}\, g^{bb'} g^{cc'} g^{dd'}
\label{eq:lmcsdecay}
\ee
where $y^a$ are the coordinates of $T^{1,1}$ and $h^{\mu\nu}$ is a
massless graviton.  Notice that the $T^{1,1}$ quantum numbers of 
the various factors are such that their
product contains a singlet (since $j,l$ are angular momentum quantum
numbers for the $SU(2)$ factors, $1\otimes 1= 0\oplus1\oplus 2$), so
the integral over $y^a$ should not vanish.  However, this vertex
involves the two different LMCS species, rather than just one of 
them and a massless mode of $A_{abcd}$, so it does not provide
phase space for the decay of the LMCS. One could possibly form a
$T^{1,1}$ singlet with other states in the vertex, but these terms are
also ruled out on kinematic grounds.

It is natural, then, to turn to the Chern-Simons part of the action
(\ref{action}) since it contains a single $A^{(4)}$ factor.  However,
the Chern-Simons action contains terms of the form
\be
	\int d^{\,5}x\, d^{\,5}y\, 
\epsilon^{abcde}\epsilon^{\alpha\beta\gamma\delta\epsilon}
	A^{(4)}_{abcd}\, \partial_e A^{(2)}_{\alpha\beta}\,
	\partial_\gamma B^{(2)}_{\delta\epsilon}\,,
\ee
with no appearance at all of the metric.  Therefore the operator
$\delta g_{ab}^{(1,1,0)}$ cannot induce decay of the LMCS through 
this term.
These two observations prove that we must consider other operators
in addition to the one invoked by ref.\ \cite{bib:aharony} for
the decay of the LMCS to proceed.

There is no {\it a priori} reason to expect that just one operator
should both give masses to the $\overline{D3}$ moduli {\it and}
mediate the decay of the LMCS.  We are free to turn on {\it any}
irrelevant operator which breaks the symmetries of $T^{1,1}$ (while
preserving other desirable symmetries) since the CY will generically
not respect the isometries of $T^{1,1}$.   

In the following subsections we will describe two decay channels
resulting from the insertion of two operators.  Before doing so, let
us mention some of the constraints which led us to these examples.
Many possibilities were ruled out due to insufficient phase space for
the decay, 4D Lorentz violation (see \cite{Elliott:2005va} for some
recent constraints), or the inability to form a $T^{1,1}$ singlet 
operator without employing relevant operators which strongly
distort the KS background in the IR region of the throat. All the
other possibilities which we found were similar to or subdominant to
those which we shall describe next.

\subsection{A SUSY-violating decay} 
\label{sec:decaychannel} 

Ideally, one might like to preserve 5D supersymmetry in the process
of modeling the effects of the CY.  This was a criterion that was
used by ref.\ \cite{bib:aharony} in identifying  $\delta
g_{ab}^{(1,1,0)}$ as the most relevant (though still irrelevant)
$T^{1,1}$-breaking operator.  However, the quantum numbers of this
operator show that by itself it is not sufficient to induce decays 
of the LMCS, $b^{(1,0,0)}$, since $(1,1,0)\otimes(1,0,0)$ does not
contain a singlet.  One way to overcome this problem is to  break
SUSY at a high scale and to solve the hierarchy problem by putting
the standard model in some other more deeply warped throat.

In analogy to the toy model example described in section \ref{simp},
an obvious choice would be to turn on a background for some KK
excitation of $A_{abcd}$, since this field
contains the LMCS, $b^{(1,0,0)}$.  We cannot use $b^{(1,0,0)}$
as a background however, because its conformal dimension is 
$\Delta =2$ (see Table \ref{tab:LMCS}), and a relevant deformation
would invalidate the KS background in the IR region.  Instead, 
one can turn on a higher mode, 
namely $A_{abcd}^{(2,1,0)}$.  This,
together with $\delta g_{ab}^{(1,1,0)}$, has
the right quantum numbers to neutralize the charge of
the LMCS $b^{(1,0,0)}$.
In the same way, $b^{(0,1,0)}$ gets its mixing from
the metric perturbation $\delta g_{ab}^{(1,1,0)}$ combined with
 $A_{abcd}^{(1,2,0)}$.  As shown in ref.\ \cite{bib:ceresole},
the conformal dimension of the operators corresponding to the
KK modes of $A_{abcd}$ (in vector multiplet I) is given by
\be
	\Delta = E_0 = \sqrt{H_0 + 4} -2\,.
\ee
Since the state $A_{abcd}^{(2,1,0)}$ has $H_0=48$, its dimension is
\be
	\Delta_{210} = 4.93 > 4\,,
\ee
and therefore turning on a small 
background for this state results in an exponentially decaying
perturbation in the throat.  
  For reference, we note that the 
 operator corresponding to the $\delta g_{ab}^{(1,1,0)}$
background used by \cite{bib:aharony} to give masses to the
$\overline{D3}$ moduli is the top component of vector multiplet I
whose conformal dimension is $\Delta = E_0 + 2$, hence
\be
	\Delta_{110} = \sqrt{28} \cong 5.29\,.
\label{D110}
\ee

\begin{figure}[!bth]\centering
\includegraphics[width=0.7\textwidth]{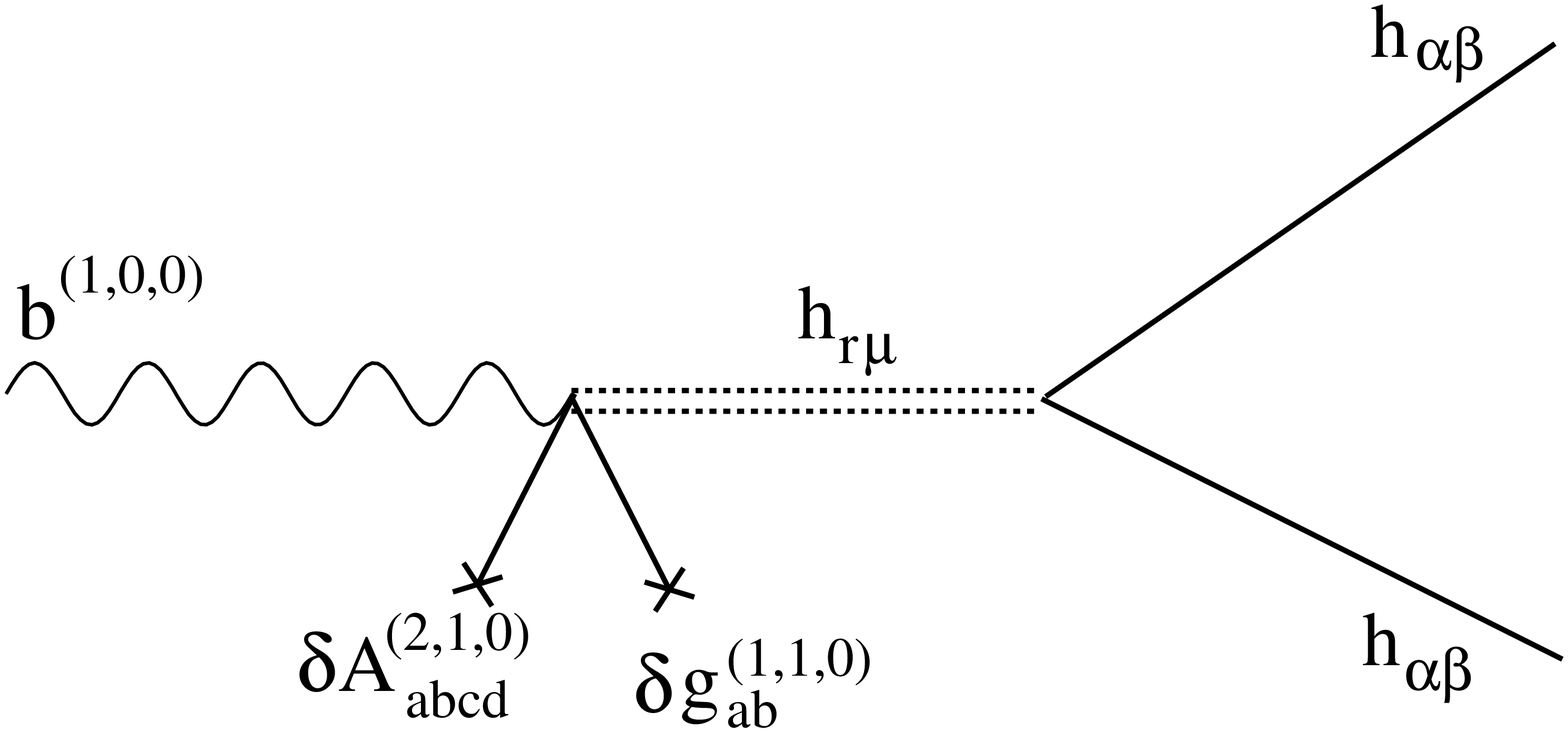}
\caption{SUSY-violating decay of the LMCS $b$ into massless gravitons
$h_{\mu\nu}$ via mixing with the radial graviphoton $h_{r\mu}$, in the
background of a tadpole for the excited state 
$\delta A_{abcd}^{(2,1,0)}$ and the metric perturbation $\delta
g_{ab}^{(1,1,0)}$ in the type IIB SUGRA model.}
\label{fig:mixing2}
\end{figure}

In this way, we obtain mixing between $b$ and the graviphoton
$h_{r\mu}$.  The zero-mode of the graviphoton is projected out
in the presence of an orbifold plane, which is present in the
compactification we consider, but we need only a massive virtual
graviphoton since it can decay into massless gravitons through the
gravitational interaction
\be
	{1\over M_p} \partial_r h^{\mu r}\partial_\mu h_{\alpha\beta}
	h^{\alpha\beta}\,.
\label{graviphoton}
\ee
The resulting decay process is shown in figure \ref{fig:mixing2}.

To compute the amplitude for the decay, we first evaluate the
integrals that give rise to mixing between $b = 
A_{a'b'c'd'}^{(1,0,0)}$ and $h^{\mu r}$ in the combined backgrounds
of $\delta g^{aa'}_{(1,1,0)}$ and 
$\partial_r \delta A_{abcd}^{(2,1,0)}$:
\be
	{\cal L}_{\rm mixing} =
	\int dr\, d^{\,5} y \sqrt{g_6}\, \sqrt{g_4}\,
	 \delta g^{aa'}_{(1,1,0)}\, h^{\mu r}\,
	\partial_r \delta A_{abcd}^{(2,1,0)}
	\partial_\mu A_{a'b'c'd'}^{(1,0,0)}\, g^{bb'} g^{cc'} g^{dd'}
\label{eq:lmcsdecay2}
\ee
For the background deformations, we use $ \delta g^{aa'}_{(1,1,0)}
\cong e^{-(\Delta_{110}-4)kr}$ and $\delta A_{abcd}^{(2,1,0)}\cong
M_{3/2}\sqrt{k} e^{-(\Delta_{210}-4)kr}$, where in the latter we make
a distinction between the size of a generic $T^{1,1}$
symmetry-breaking tadpole, $M_s$, and the scale $M_{3/2}$ at which
SUSY is broken.  Moreover the radial wavefunction for the LMCS $b$ is
approximated by  $R_{(1,0,0)}\sim  w^{1+\nu_b}\sqrt{k}\, e^{(2+\nu_b)kr}$ with
$\nu_b = \sqrt{4 + m^2_b} =  0$ (see discussion above eq.\
(\ref{effint2})).  For the vector $h_{r\mu}$ on the other hand, the
radial wave function is approximated by $R_{(0,0,0)} \sim  w^{1+\nu_h}\sqrt{k}\,
e^{(1+\nu_h)kr}$ and $\nu_h = \sqrt{1+m^2_{5D}/k^2}$, as shown in appendix \ref{vectors}.  Recalling that
$B_\mu$ is the 4D field corresponding to $h_{r\mu}$, to find the
largest amplitude, we need the smallest 5D mass eigenvalue for $B_\mu$
or $\phi_\mu$ (as explained in ref.\ \cite{bib:ceresole}, the mass
eigenstates are mixtures of $\phi_\mu$ and $B_\mu$); this appears in 
gravitino multiplet I, with $H_0^-(0,0,0) = -3/4$ and $m^2/k^2 = 
7 + H_0^- - 4\sqrt{4+H_0^-}\cong -0.96$, giving $\nu_h \cong 0.20$. 
 Remembering $\sqrt{g_4} = e^{-4kr}$ and $h^{\mu r} = e^{2kr} h_{\mu r}$, the radial
integral becomes
\be
	w^{2+\nu_h} M_{3/2} k^2 \int dr \, e^{(9+\nu_h-\Delta_{110}-\Delta_{210})kr}
 \sim w^{2.2} M_{3/2} k\,.
\label{eq:pint}
\ee
Combining this with the interaction (\ref{graviphoton}), we get
the effective decay vertex
\be
	{\cal L}_{\rm decay} \sim 
	 w^{2.2}\, {M_{3/2} k\over M_p m^2_h}\,
	\partial_\mu b\, h^{\alpha\beta}\, \partial^\mu h_{\alpha\beta}
\label{eq:TdecayJ}
\ee
where $m_h\sim w k$ is the mass of the exchanged graviphoton.
Taking $m_b$ to be also of this order, we find that the decay rate
is 
\be
	\Gamma\sim w^{3.4} {M_{3/2}^2 k\over M_{p}^2}
\label{mainres}
\ee
and the decay temperature is of
order $T\sim w^{1.7} M_{3/2}(k/M_p)^{1/2}$.  
Assuming the inflationary scale is of order
$10^{14}$ GeV \cite{KKLMMT} and $k\sim M_p$, then $w\sim 10^{-4}$,
 and $T\sim 10^{-2} M_{3/2}$.  For
electroweak baryogenesis to work, the relics should decay before
$T\sim 100$ GeV, which leads to a constraint on the SUSY-breaking
scale,
\be
	M_{3/2} > 100\, w^{-1.7} {\rm \ GeV}\cong 6\times 10^8 {\rm \ GeV}\,.
\ee
It was pointed out in ref.\ \cite{KL1} that the gravitino mass is
 larger than this bound in the  KKLT construction 
\cite{KKLT} on which the warped brane-antibrane inflation scenario
is based.  Therefore we conclude that there is not a problem with
heavy KK relics in the $T^{1,1}$ background.

\subsection{SUSY-preserving decay}
\label{sec:susydecay}

\begin{figure}[!bth]\centering
\includegraphics[width=0.7\textwidth]{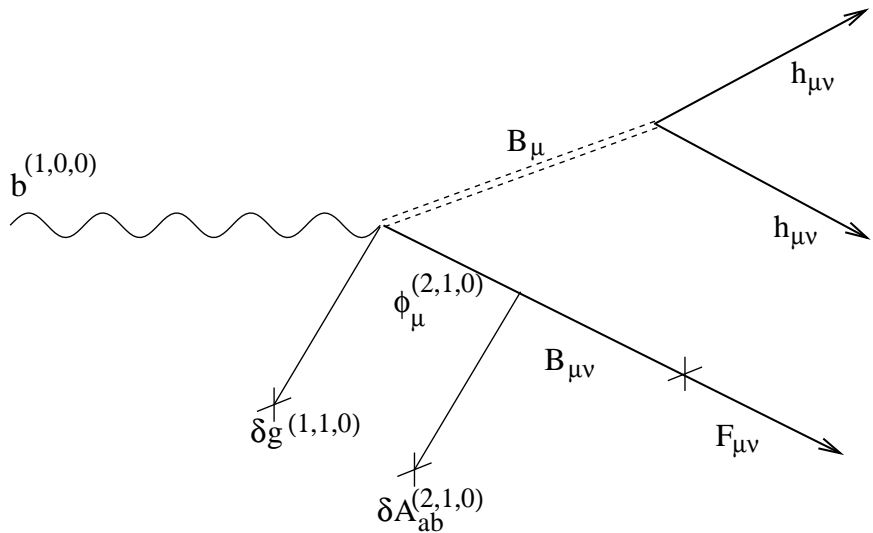}
\caption{SUSY decay of the LMCS $b$ into massless gravitons
$h_{\mu\nu}$ and gauge boson in the
background 
$\delta A_{ab}^{(2,1,0)}$ and the metric perturbation $\delta
g_{ab}^{(1,1,0)}$.}
\label{fig:susydecay}
\end{figure}

It is also possible to find decay channels which do not require
SUSY-breaking deformations of the background.  Turning on the
same SUSY background $\delta g^{(1,1,0)}$ as previously, 
 the kinetic term for the 4-form gauge field contains the terms
\be
	\delta g^{(1,1,0)} h^{\mu d}
	\partial_r A^{(1,0,0)}_{abcd}
	\partial_r A^{(2,1,0)}_{abc\mu}   
	\ \sim\ k^2 B^\mu\, b^{(1,0,0)}\,
	\phi^{(2,1,0)}_\mu\,.
\ee
The vector fields $B^\mu$ and $\phi_\mu$ are both taken as massive.  The 
former can decay into massless gravitons similarly to the graviphoton
$h_{\mu r}$, through the interaction
\be
	\partial^a h_{\mu a} h_{\alpha\beta}\partial^\mu 
	h^{\alpha\beta}
\ee
contained in the 10D Ricci scalar.
Moreover, we can generate mixing between $\phi_\mu$ and a massless
particle by turning on an additional SUSY deformation of the
2-form gauge field, $\delta A_{ab}^{(2,1,0)}$ and using it in the 
Chern-Simons (CS) action $A_4\wedge F_3\wedge H_3$,
\be
\epsilon^{abcder\mu\nu\rho\sigma}	
A^{(2,1,0)}_{abc\mu}\, \delta A_{de,r}^{(2,1,0)}\,B_{\nu\rho,\sigma}
\ \sim \ \epsilon^{\mu\nu\rho\sigma}\phi_\mu^{(2,1,0)} B_{\nu\rho,\sigma}\,.
\ee
The massless mode of the Kalb-Ramond field is projected out of the
spectrum similarly to the massless graviphoton, but if we assume there
is a stack of D-branes in the throat, then $B_{\nu\rho}$ mixes with
a massless gauge field on the brane stack via the DBI action.  The
complete process is illustrated in figure \ref{fig:susydecay}.
To estimate the amplitude, we consider the $b B_\mu\phi^\mu$ vertex
and the $\phi_\mu B_{\mu\nu}$ mixing amplitudes separately.  The
effective coupling of the former is given by
\be
	{\cal L}_{bB\phi} \cong	
b B_\mu\phi^\mu\, k^2\,w^{\sum_i(1+\nu_i)} \int dr e^{-2kr + (4+\sum_i\nu_i)kr
	+ (4-\Delta_{110})kr}
\ee
where $\nu_i$ refer to the fields in external lines,
$i=b,B_\mu,\phi_\mu$ and
the $e^{-2kr}$ factor comes from $\sqrt{g_4} g^{\mu\nu}$.  The
quantities $\nu_i$ and the conformal dimension $\Delta_{110}$ can be
deduced from ref.\ \cite{bib:ceresole}.  We have already noted that
$\nu_b=0$ and $\nu_B=0.2$.  Appendix C of ref.\ \cite{bib:ceresole} gives
$m^2_{5D}/k^2 = 60-6\sqrt{52} \cong 16.73$ for $\phi_\mu^{(2,1,0)}$
in vector multiplet I.
Thus $\nu_\phi = \sqrt{1 + m^2_{5D}/k^2} \cong 4.21$.  The conformal dimension
$\Delta_{110}\cong 5.29$ of $\delta g^{(1,1,0)}$ was already given in
(\ref{D110}).   We then obtain
\be
	{\cal L}_{bB\phi} \cong k\, w^{\Delta_{110}-3}\, 	
b B_\mu\phi^\mu\
\ee 
which is suppressed by $w^{2.29}$. This conclusion is completely 
insensitive to the values of $\nu_i$,  because the integral is
dominated by the IR contributions, and the factors of $w^{\nu_i}$
therefore cancel between the normalization and the integral.

Similarly, using the approximation for the $B_{\mu\nu}$ radial
wave function $w^{1+\nu}\sqrt{k} e^{\nu kr}$ 
(see appendix \ref{kalb-ramond} and above eq.\ (\ref{effint2})) the mixing amplitude is
\be
{\cal L}_{B\phi} \cong 	\epsilon^{\mu\nu\rho\sigma}\phi_\mu B_{\nu\rho,\sigma}
	 \, w^{\sum_i(1+\nu_i)} \int dr e^{-4kr + (3+\sum_i\nu_i)kr
	+ (4-\Delta_{210})kr}	
\ee
where $i=\phi_\mu, B_{\mu\nu}$, and $e^{-4kr} = \sqrt{g_4}$ is not accompanied by any
inverse metric factors in the CS action.   We already have 
$\nu_{\phi_\mu} = 4.21$ from above; for $\nu_{B_{\mu\nu}}$ we refer
to the states $a_{\mu\nu}$ in  ref.\ \cite{bib:ceresole}.  The lowest
value of $m_{5D}/k$ for $a_{\mu\nu}^{(0,0,0)}$ is found
to be $2 - \sqrt{4-3/4} = 0.197$ in gravitino multiplet 
I.\footnote{Masses of two-form states in appendix C of ref.\ 
\cite{bib:ceresole} are given without being squared, and without
regard to the overall sign}\   
From section 
\ref{kalb-ramond} we thus see  that 
$\nu_{B_{\mu\nu}} = \sqrt{m^2_{5D}}/k \cong 0.2$.
On the other hand, the lowest irrelevant value of $\Delta_{210}$
for the deformation $\delta A_{ab}^{(2,1,0)}$ comes from the entry
for $a$ in gravitino multiplet I or III with $m^2 = H_0^- + 4 -
4\sqrt{H_0^-+4} = 22.6$, giving $\Delta_{210}=
2 + \sqrt{26.6} = 7.16$.  Then we find that the $r$ integral is
marginally dominated by the IR contribution, hence
the result is insensitive to $\sum_i\nu_i$
as far as powers of the warp factor are concerned.  We obtain
$w^{\Delta_{210}-1}$:
\be
{\cal L}_{B\phi} \cong k w^{6.16}\epsilon^{\mu\nu\rho\sigma}\phi_\mu B_{\nu\rho,\sigma}\,.
\ee

Finally, we can construct an effective vertex by integrating out the
intermediate $\phi_\mu$, $B_\mu$ and $B_{\mu\nu}$ states.  For the 
latter, we assume that there is a brane stack at $r=r_0$ in the
bottom of the throat, with DBI action
\be
	\int d^{\,4}x\sqrt{\det(g_{\mu\nu} + B_{\mu\nu} + F_{\mu\nu})}
\ee
which gives rise to the mixing term $\eta^{\alpha\mu}
\eta^{\beta\nu}B_{\alpha\beta}F_{\mu\nu} = B\cdot F$.  The important
point is that $\sqrt{g}g^{\alpha\mu}g^{\beta\nu}$ is independent
of the warp factor, so no extra powers of $w$ arise due to the
$B$-$F$ mixing. The effective interaction
becomes
\be
{\cal L}_{bhhF} = w^{8.45} {k^4\over{M_p (wk)^6} }\,b\, h_{\alpha\beta}\partial_\mu 
h^{\alpha\beta}\, \epsilon^{\mu\nu\rho\sigma}F_{\nu\rho,\sigma}\,,
\ee
where the inverse powers of $wk$ come from the propagators for the
intermediate states $B_\mu$, $\phi_\mu$ and $B_{\mu\nu}$, since
all have masses of order $wk$.  
Furthermore each derivative (of which there are three) will be of
order $m_b\sim wk$ in the decay amplitude, leading to ${\cal L}_{bhhF}
\sim w^{5.45}$.  Squaring this, and taking account of the phase space
going like $m_b\sim wk$, we obtain for the decay rate
\be
	\Gamma \cong w^{11.9} {k^3\over M_p^2}
\ee
The decay temperature is $T\sim w^6\sqrt{k^3/M_p}$.  
For $w\cong 10^{-4}$ and $k\sim M_p$, the relic will decay at $T\sim
10^4$ eV, which is too late to be consistent with nucleosythesis.
We have not been able to find a faster decay channel consistent with
a SUSY-preserving background, hence the SUSY-violating decay of 
the previous section is the relevant one.

\subsection{More general bounds} 
\label{sec:general}

\begin{figure}[t]
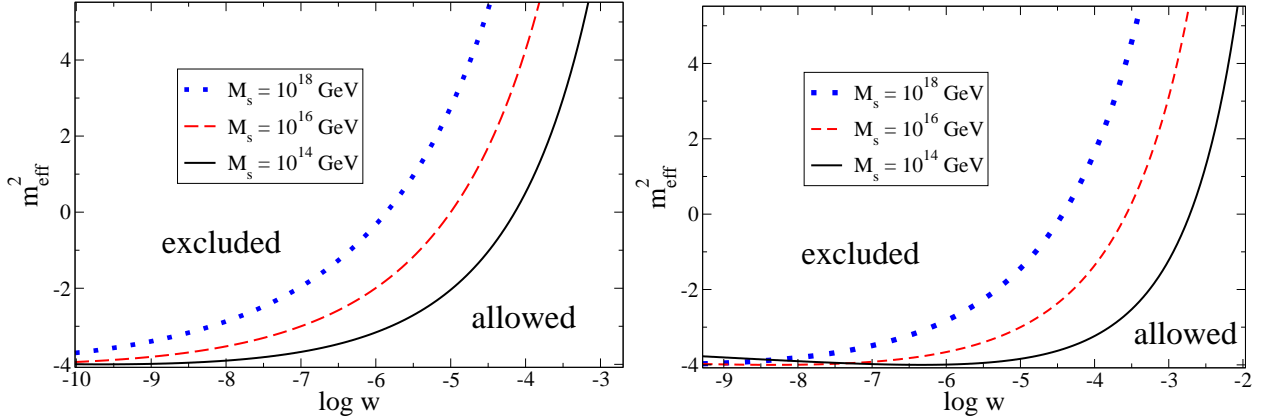
\centering
\centerline{\includegraphics[width=0.5\textwidth]{bbnlimit.eps}
\includegraphics[width=0.5\textwidth]{baryolimit.eps}}
\caption{Regions of allowed and excluded parameter space 
(see eq.\ (\ref{meff})) from requiring 
that the LMCS decay before the onset of BBN (left) and before
electroweak baryogenesis (right). The different contours
 correspond to different values of the string scale $M_s$.
$m^2_{\rm eff}$ is in units of $k^2$.}
\label{fig:m5_vs_T}
\end{figure}

In other warped background geometries, for example using the 
manifold $Y^{p,q}$ instead of $T^{1,1}$ \cite{gaunt}, the low-energy
spectrum may be different, resulting in different masses and decay
chains for the LMCS.  Regardless of these details, however, we can
expect some basic similarities to our previous analysis. The kinetic
term will always enable the process shown in fig.\ 
\ref{fig:mixing2},  with the insertion of  background corrections 
corresponding to an angularly excited state of the LMCS and of the
internal metric absorbing the charge of the LMCS.  

As an example of how our results might generalize to other
backgrounds, we will consider the masses (conformal dimensions) of
the LMCS and the lightest graviphoton to be free parameters, 
and let $\nu = \nu_b + \nu_h$.  We define an effective 5D mass,
\be
	m^2_{\rm eff} = k^2(\nu^2 - 4)
\label{meff}
\ee
which approximately matches the mass of the LMCS in the $T^{1,1}$
background, since $\nu_h\cong 0.2$ is small in that background.
Assuming that the radial integral for the mixing between $b$ and
$h_{r\mu}$ is still IR dominated (hence sensitive to $\nu$ rather
than the $\Delta$'s of the background insertions), the decay rate
for the LMCS takes the form
\be
\Gamma\simeq {M_s^3\over M_p^2} w^{2\nu+3}
\label{eq:drate}
\ee
Here we ignore the distinction between the scales $M_{3/2}$ and $k$,
assuming that $M_s$ is the only relevant scale aside from
$M_p$.  By demanding that the relics have decayed before
nucleosynthesis, occurring at the temperature
$T_{BBN}\sim 1$ MeV, we obtain the constraint
\bea
\sqrt{4+m_{\rm eff}^2}\le 
\ls \frac{\log\lb M_p T_{BBN}^2/M_s^3\rb}{2\log\lb w\rb}-\frac32\rs\,.
\eea
A similar but more stringent constraint arises from requiring the
relics to decay before baryogenesis, which can reasonably happen no
later than the electroweak scale, $T_{EW} \sim$ 100 GeV.
Fig.~\ref{fig:m5_vs_T} shows these bounds in the parameter space
of $m^2_{\rm eff}$ of the LMCS versus $w$ for several different
values of $M_s$. A background falling within the excluded region of
Fig.~\ref{fig:m5_vs_T} will spoil BBN or baryogenesis by
modifying the standard Hubble rate or though its late
decays.

\subsection{Decays versus annihilation}
\label{comparison}

Complementary constraints may be obtained by considering 
annihilation processes of the LMCS instead of their decays.
The former are efficient for small warp factors, while the latter
dominate at larger values of $w$.  
From
eq.\ (\ref{eq:drate}), we see that the LMCS decays by a temperature
$T_d \sim \sqrt{\Gamma_d/M_{pl}}$ provided the warping satisfies
\bea
w&>&\lb\frac{T_d^2 M_{pl}}{M_s^3}\rb^{1/(2\nu+3)}\,.
\eea
(where the inequality indicates that larger $w$ results in an earlier
decay).  On the other hand, the
work of ref.~\cite{Chen:2006ni} (in particular, eq.\ (5.11))
showed that in order to have fast enough annihilations, the warp factor
must satisfy 
\bea
w< \ls {g\rho_{\rm eq}\over M_p^4}\lb\frac{L}{R}\rb^{24}\rs^{1/8}\lsim 10^{-8}\,,
\label{west}
\eea
where $g\sim 100$ is the number of relativistic degrees of freedom, 
$L/R$ is the ratio of the bulk Calabi-Yau size to the AdS curvature
scale, and $\rho_{\rm eq}\simeq 1\rm{eV}$ is the energy density at the
time of radiation-matter equality.
This constraint arises by demanding that KK relics do not dominate
the energy density of the universe at this epoch.
It implies an upper bound on the warp factor
because the annihilation rate relies on the radial overlap of two
massive modes; however, massive modes are peaked more in longer throats, so
annihilation is most efficient in heavily warped throats.  

To get $w$ as large as $10^{-8}$ in (\ref{west}), it was necessary to
take $L\sim 100 R$.  One would typically like $w\sim 10^{-4}$ to
obtain the scale of inflation in the throat which is found
from the COBE normalization for brane-antibrane inflation.  This poses
less of a challenge for the decay scenario.  Assuming
$M_s\sim M_p$ and $\nu=0$, and demanding the LMCS to decay before the
electroweak scale $T\simeq 100\, \rm{GeV}$ gives the constraint
\bea
w>\lb\frac{T_{EW}^2}{M_{pl}^2}\rb^{1/3}\simeq 10^{-11} 
\eea 

We see that the single throat scenario has
complementary constraints from decays and annihilations for avoiding the
KK relic problem.  A concern would arise if these
constraints did not have an overlapping region.
For instance, a background whose LMCS has $m_{5d}^2=-2$ (in AdS
units), or $\nu=\sqrt{2}$, requires $w>10^{-5}$ in order to decay
before the EW timescale. In this case, backgrounds whose warping lay
in the range $10^{-8}<w<10^{-5}$ will result in a relic density of
massive particles, since neither annihilation nor decay is efficient
enough.

\section{Discussion}
\label{sec:discussions}

In this paper we have studied the problem of potentially long-lived
KK relics localized in a KS throat following brane-antibrane
inflation.  We found that accidental symmetries prevent the lightest
mass charged state (LMCS) from decaying, even if the angular
isometries giving rise to the conserved KK quantum numbers are broken
by the Calabi-Yau manifold to which the throat is glued.  In
ref.~\cite{bib:kofmanyi}, it was assumed the isometry-breaking
effects of the operator $\delta g_{ab}^{(1,1,0)}$ provided a decay
channel for the LMCS; however, this state either doesn't produce a
$T^{1,1}$ singlet and  so the angular integral vanishes for the
effective 4D interaction, or it doesn't couple to states with light
decay products and so there is no phase space for the process to
proceed. Instead, for the KS background one must consider
deformations of the throat geometry which break not only the
isometries but also supersymmetry; additionally, we stress that any 
background correction used to mediate the decay must correspond to 
an irrelevant operator, otherwise the background solution is disrupted. 
A term satisfying these constraints was used to derive the viable 
interaction in eq.~(\ref{eq:TdecayJ}).     

In that estimate, we omitted factors of order unity, {\it e.g.},
$1/(\Delta_{110}+\Delta_{210}-8)$, which contain the dependence on
the dimension of the symmetry-breaking operators. The important
point is that even if the $T^{1,1}$-breaking operators have a very
high dimension,  this only mildly suppresses the strength of the
interaction since one integrates the radial profiles over the length
of the throat. The 4D effective coupling receives UV and IR
contributions, with suppression in the IR determined by the
dimension of the symmetry breaking in the CY, and suppression in the
UV controlled by the behaviour of the radial wave function of the 
LMCS (see Appendix \ref{app:app2} for more details). We find that
the dependence of the decay amplitude  on the warp factor is
$w^{2+\nu_b + \nu_h}$, coming from the LMCS and graviphoton radial
wave functions. It depends on the 5D masses of the particles,   in
units of the AdS curvature, through $\nu = \sqrt{4+m^2_{5D}}$. This
leads to a parametrically different rate of decay in terms of powers
of the warp factor than the guess which was based upon the dimension
of the symmetry-breaking operator.  

We find nontrivial constraints on the parameters by requiring the
heavy KK relics to decay before nucleosynthesis, and more
stringently, before electroweak baryogenesis; the SUSY-breaking
scale should exceed $\sim 10^9$ GeV if $w\sim 10^{-4}$ as needed for
brane inflation to give the right amplitude of density
perturbations.  For other warped backgrounds, in which the states
 might have different masses, one can also constrain the effective
LMCS mass versus the warp factor, as we have shown in figure
\ref{fig:m5_vs_T}.

\bigskip {\bf Acknowledgements.} The authors would like to thank R.\
Brandenberger, C.\ Burgess, X.\ Chen, K.~Dasgupta, O.~Aharony,
 A.~Ceresole, A.~Frey,  B.v.\ Harling and S.~Kachru  for many
useful discussions.   A.B.\ and J.C,\ are supported by NSERC (Canada) and FQRNT
(Qu\'ebec).  H.S.\ is supported by the EU under MRTN contract
MRTN-CT-2004-005104 and by PPARC under rolling grant PP/D0744X/1.

\appendix

\section{Dimensional Reduction}
\label{app:app2}

In this appendix we give details of the dimensional reduction 
from the 10D SUGRA action to 4D for scalar and vector fields,
in the background of operators which perturb the KS solution for
the warped throat.

\subsection{Scalar fields}

In the KS background the LMCS comes from the RR
4-form polarized along the $T^{1,1}$, which a 4D observer sees as a
scalar field. We use this field as the eponymous example for scalar
fields in the AdS; other 5D scalars will differ in their bulk mass
spectrum, but their behaviour along the internal directions will be
similar.  

Rewriting the IIB supergravity action with terms involving
$A_{(4)}$ gives
\bea
S_{IIB}(A_{(4)})&=&\frac{1}{2\kappa_0^2}\int
  d^{10}x\sqrt{-G}\ls-\frac{1}{240}\lb
  F^{(5)}\rb^2\rs+\frac{1}{2\kappa_0^2}\int A^{(4)}\wedge
  F^{(3)}\wedge H^{(3)}\,.\nn\\
\label{eq:4formact}
\eea
In this expression $\kappa_0$ is related to the 10D Newton constant,
$F^{(n+1)}=dA^{(n)}$ is the field strength of the RR 4-form, and
$H^{(3)}$ is the field strength of the NS 2-form $B^{(2)}$. To extract
the conditions necessary to obtain a canonically normalized 4D scalar
field for the LMCS we expand about background values of the metric,
employ the harmonic expansion
\bea
A_{abcd}(x,y)&=&\sum_{m,\set{\nu}}\tilde b(x^\mu)R_{m\set{\nu}}(r){\epsilon_{abcd}}^eD_eY_\set{\nu}(y)\,,
\eea
and isolate the 4D kinetic term in~(\ref{eq:4formact})
\bea
\Rightarrow S_{IIB}(A_{(4)})&=&\frac{-1}{480\kappa_0^2}\int d^{4}x\,dr\,d^5y
\sqrt{-|G_{AdS_5}|}\sqrt{|G_{T^{1,1}}|}
\,G^{a_2b_2}G^{a_3b_3}G^{a_4b_4}G^{a_5b_5}\nn\\&&
\left.\hspace{.71cm}
\times G^{\mu\nu}\del_{\mu}A_{a_2a_3a_4a_5}\del_{\nu}A_{b_2b_3b_4b_5}
\rs\,.\label{eq:ints}
\eea
To reduce this to a standard kinetic term in 4D for the LMCS, we
require the following:
\bea
&&\frac{1}{\sqrt{|G_{T^{1,1}}|}}\del_a\ls\sqrt{|G_{T^{1,1}}|}\,
{\epsilon_{bcde}}^f{\epsilon^{bcdeg}}
\del^a\lb\mathcal{D}_f\mathcal{D}_gY_\set{\nu}\rb \rs=-H_\set{\nu} Y_\set{\nu}\\
\label{a4}
&&\Box_r R_n-m_{5D}^2e^{-4kr}R_n=-m_n^2e^{-2kr}R_n
\label{a5}
\\
&&\frac{-1}{240}\int d^5y
\sqrt{|G_{T^{1,1}}|}\,{\epsilon_{bcde}}^f 
{\epsilon^{bcdeg}}\mathcal{D}_fY_\set{\nu}\mathcal{D}_gY_\set{\mu}
=\delta_{\{\mu\},\{\nu\}}  
\label{a6} 
\\
&&\int dr\,e^{-2kr}\,R_m
R_n=\delta_{m,n}\,. 
\label{a7}
\eea
Eqs. (\ref{a4},\ref{a5}) are the equations of motion for
$T^{1,1}$ and radial wave functions respectively, with 4D
mass $m_n$, while
(\ref{a6},\ref{a7}) 
are the orthonormality conditions. $H_\set{\nu}$ is the
eigenvalue of the Laplacian on $T^{1,1}$ 
for the eigenfunction $Y_\set{\nu}$, given by
\be
H_\set{j,l,r}=6\ls 
j(j+1)+l(l+1)-r^2/8\rs
\ee
From Table C (Vector Multiplet I) of ref.\ \cite{bib:ceresole}, 
$H_{\set{\nu}}$ determines the 5D mass of the LMCS by
\be
m_{5D}^2=H_\set{\nu}+16-8\sqrt{H_\set{\nu}+4}\,.
\ee
 $\set{\nu}$ is the
set of quantum numbers specifying the representation of 
 $SU(2)\times SU(2)/U(1)$ ($T^{1,1}$). 

In analogy to the RS model \cite{RS,bib:goldwise}, the solutions for
the  radial wave functions have the form
\bea
R_n(r)&=&\frac{\sqrt{k}\,we^{2kr}}{J_\nu(x_n)}\ls J_\nu\lb
x_nwe^{kr}\rb
+b_{n\nu}Y_\nu\lb x_nwe^{kr}\rb\rs\,,\,\nu=\sqrt{4+H_\set{\nu}}\,,
\label{eq:R}
\eea
where $w=e^{-kr_c}$ is the warp factor and $x_n\equiv {m_n
e^{kr_c}}/{k}\sim\mathcal{O}(1)$.   Thus $m_n\sim w k$, which is
of order the scale of inflation in the present application, while
$k$ is of order the Planck scale.   This
solution assumes the bulk CY is not exponentially large
compared to the length of the
throat, $r_c$, so that the normalization factor is determined by
integrating only over the throat itself.\footnote{ 
Ref.~\cite{bib:hassan} examined the effect of a large bulk on the
4D mass spectrum and the transition to the 
ADD scenario~\cite{bib:add} in the limit that the bulk becomes 
much larger than $(kw)^{-1}$.  
The effect of a
large bulk to suppress tunneling between throats 
was also explored in ref.~\cite{bib:hhn}.  The work
presented here is concerned mainly with reheating in one throat and
envisions possible tunneling to a SM throat later on.} 

It is worth noting the asymptotic limits of the Bessel
functions, since these determine the UV and IR behaviour of the 
radial
wavefunctions and play an important role in determining the 4D effective
coupling. In the UV region, $r\ll1$ and $x_nw e^{kr}\simeq x_nw\ll1$, and we
employ the small-argument expansion  
\bea
R_n(r)\simeq\frac{\sqrt{k}we^{2kr}}{J_\nu(x_n)}\ls\lb
x_nwe^{kr}\rb^\nu+b_{n\nu}\lb x_nwe^{kr}\rb^{-\nu}\rs\,.
\eea
Taking $Z_2$ orbifold boundary conditions as in the RS model 
fixes the coefficient $b_{n\nu}$ to be
\bea
b_{n\nu}=-\frac{2J_\nu(x_nw)+x_nwJ_\nu^\prime(x_nw)}{2Y_\nu(x_nw)+x_nwY^\prime_\nu(x_nw)}\simeq -(x_nw)^{2\nu}
\eea
so that both solutions are of the same order ($R_n(0)\simeq\sqrt{k}\,w^{\nu+1}$) in
the UV.

In the IR, $r\simeq r_c$, such that $x_nw e^{kr}\simeq1$ and
\bea
R_n(r)\simeq\frac{\sqrt{k}we^{2kr}}{J_\nu(x_n)}\ls\sqrt{\frac{2}{x_nwe^{kr}}}\cos\lb
x_nwe^{kr}-\lb\nu+\frac12\rb\frac{\pi}{2}\rb \rs\sim\sqrt{kw}\,e^{3/2kr}\,.
\eea
In this case the coefficient $b_{n\nu}$ makes the $Y_\nu$ solution
negligible, so the IR behaviour is dominated by the $J_\nu$ solution. At
the bottom of the throat $r=r_c$ and $R_n(r_c)\sim \sqrt{\frac{k}{w}}$,
so the wavefunction is exponentially peaked in the IR.

\subsection{Vector fields}
\label{vectors}

To give an example of DR for a vector field in a warped space
\cite{bib:rizzo}, we consider the 4-form with one index polarized
along the AdS, while the other three are along the $T^{1,1}$:
\bea
A_{\mu abc}(x,r)&=&\sum_\set{\nu}\phi_\mu^\set{\nu}(x)Y^\set{\nu}_{abc}(y)\,.
\eea
(in this section Greek indices include $r$ in addition to the
noncompact directions).
The kinetic term is
\bea
&&
S_{\rm kin}(\phi_r)=-\frac{1}{480\kappa_0^2}
\int d^{10}x\sqrt{-G}\,
\del_{[\alpha}A_{\beta]cde}\del^{[\alpha}A^{\beta]cde}
\label{a12}\\
&&
=-\frac{1}{480\kappa_0^2}\sum_{\set{\mu}\,\set{\nu}}
\int d^{\,5}x\sqrt{-|G_{AdS_5}|}\, 
\del_{[\alpha}\phi_{\beta]}^\set{\mu}\del^{[\alpha}\phi^{\beta]}_\set{\nu}
\int d^{\,5}y\sqrt{|G_{T^{1,1}}|}\,
Y^\set{\mu}_{cde}Y_\set{\nu}^{cde}\,.
\eea
We therefore normalize
\bea
&&\frac{1}{120}\int d^{\,5}y\sqrt{|G_{T^{1,1}}|}
 \,Y^\set{\nu}_{cde}Y_\set{\mu}^{cde}
=\delta_{\{\mu\},\{\nu\}}
\eea
which gives the 5D action for the vector field $\phi_\gamma$:
\bea
S_{\rm kin}(\phi_r)&=&-\frac{1}{4\kappa_0^2}\int d^{\,5}x
\sqrt{-|G_{AdS_5}|}\,
 ( \del_{[\alpha}\phi_{\beta]}
\del^{[\alpha}\phi^{\beta]} + m^2_{5D}\phi_\alpha\phi^\alpha)
\,.
\label{a15}
\eea
Here we have allowed for a 5D mass, which does not come from the
terms in (\ref{a12}), but which would generically arise from 
derivatives along the $T^{1,1}$ directions.

The equation of motion arising from (\ref{a15}) is
\be
\partial^\alpha F_{\alpha\beta} -m_{5D}^2 \phi_\beta = 0
\ee
where $F_{\beta\delta}=\partial_{[\beta}\phi_{\delta]}$.
This leads to an equation for the radial wave function whose solution
has the form
\be
R_{1-form}(r)\simeq\sqrt{k}\,we^{kr}\ls J_\nu\lb x_nw e^{kr}\rb 
+b_{n\nu}Y_\nu\lb x_nwe^{ky}\rb\rs,\,\nu=\sqrt{1+m^2_{5D}/k^2}\,.
\label{vector_soln}
\ee

\subsection{Antisymmetric tensor field}
\label{kalb-ramond}
The equation of motion for the rank-2 antisymmetric tensor field is:

\be
\frac{1}{\sqrt{-G}}\partial_{A}\left[\sqrt{-G}F^{ABC}\right] + M^2 B^{BC} = 0
\ee
where $M$ is the 5-dimensional mass given by the non-trivial $T^{1,1}$ quantum
numbers of the field. 
We want to write the equation in terms of $B_{\mu\nu}$ so we rewrite the 
above equation as:
\be
\frac{1}{\sqrt{-G}}\partial_{A}\left[\sqrt{-G}g^{AL}g^{BM}g^{CN}F_{LMN}\right] + 
M^{2}g^{BM}g^{CN} B_{MN} = 0
\ee
where the metric is the RS metric:
\be
ds^2 = e^{-2kr}\eta_{\mu\nu}dx^{\mu}dx^{\nu}+dr^{2}
\ee
and the field strength is:
\be
F_{LMN}=\frac{1}{6}\left[
  \partial_{L}B_{MN} -
  \partial_{L}B_{NM} +
  \partial_{M}B_{NL} -
  \partial_{M}B_{LN} +
  \partial_{N}B_{LM} -
  \partial_{N}B_{ML} 
\right]
\ee
For $B$ and $C$ corresponding to polarizations 
along the large 4 dimensions we get:
\baray
&& e^{4kr}\partial_{\lambda}\left[e^{-4kr}e^{6kr}\left(
    2\partial^{\lambda}B_{\mu\nu} + 
    2\partial_{\mu}B^{\lambda}_{~\nu}+ 
    2\partial_{\nu}B^{~\lambda}_{\mu}
  \right)\right] + \nonumber \\
&& e^{4kr}\partial_{r}\left[e^{-4kr}e^{4kr}\left(
    2\partial^{r}B_{\mu\nu} + 
    2\partial_{\mu}B^{r}_{~\nu}+ 
    2\partial_{\nu}B^{~r}_{\mu}
  \right)\right] +\nonumber \\
&&  e^{4kr}M^{2}B_{\mu\nu} = 0
\earay
To proceed we will need to fix the gauge, 
i.e. we need a condition of the trpe:
\be
g^{MN}\partial_{M}B_{N\nu} = 
e^{2kr}\partial_{\lambda}B^{\lambda}_{~\nu}+
\partial_{r}B^{r}_{~\nu} = 0
\ee
The equation of motion simplifies to:
\be
e^{2kr}\Box_{4}B_{\mu\nu}  -
\partial_{r}^{2}B_{\mu\nu} + 
M^{2}B_{\mu\nu} = 0
\ee
If we now decompose:
\be
B_{\mu\nu}\left(x,r\right) = 
B_{\mu\nu}\left(x\right)\,R\left(r\right)
\ee
we obtain for $\chi\left(r\right)$ the equation:
\be
-\partial_{r}^{2}R\left(r\right)+
M^{2}R\left(r\right)-
e^{2kr}m^{2}R\left(r\right) = 0
\ee
which has a solution of the type:
\be
R(r) =   J_\nu(x_n w e^{kr}) + 
b_{n\nu} Y_\nu(x_n w e^{kr}),\quad \nu={m_5\over k}
\ee
This shows that for the rank-2 antisrmmetric tensor, 
$\nu = {m_5}/{k}$.

\subsection{Background deformation by source in UV}
\label{sec:tadpole}

To model the effect of gluing a Calabi-Yau onto the throat, and
the ensuing deformation, we can
introduce a source term localized in the UV, which is 
the  CY region. As an example we consider a background
perturbation of the 4-form, with $T^{1,1}$ quantum numbers
$\set{\nu}$, and  coupled to a source of strength
$S_\set{\nu}^{abcd} \sim M_s^4\tilde\phi$, localized in the UV region, which we take to be
at $r=0$.  The action for the source term is 
\bea
S_\set{\nu}&=&\int d^{10}x\sqrt{-G}\, S_\set{\nu}^{abcd}A_{abcd}
\nn\\
&=&\int d^{10}x \sqrt{-G}\underbrace{\tilde\phi M_s^4\delta(r)
{\epsilon^{abcdf}}
\mathcal{D}_fY_\set{\nu}}_{\rm source}\ \cdot\ 
\underbrace{\sum_\set{\nu^\prime}b_\set{\nu^\prime}(x){\epsilon_{abcd}}^g
\mathcal{D}_gY_\set{\nu^\prime}(y)}_{\rm field}\,.
\label{eq:source}
\eea

Upon variation of the field, the source appears in the equation of 
motion for the radial wave function of the field,  
\bea
\Box_rR_{0,\set{\nu}}-m_{5D}^2e^{-4kr}R_{0,\set{\nu}}&=&\tilde\phi
\delta(r)\,.
\eea
where the subscript $0$ indicates that the 4D mass is zero, 
since by Lorentz invariance this perturbation of the background
cannot depend on the large dimensions.
There are two particular solutions, of which one is
subdominant over the entire throat, so we can approximate
\bea
R_{0,\set{\nu}}(r)&\cong&\sqrt{k}\tilde\phi
e^{(2-\nu)kr}=\sqrt{k}\,\tilde\phi\,e^{(4-\Delta )kr}\,
\label{eq:scalar_corr}
\eea
where $\nu=\sqrt{4+m_{5D}^2}$ and  $\Delta=\nu+2$ is the conformal
dimension of the operator. In order that the KS  solution not be
strongly deformed in the IR,  we demand that $\Delta>4$.\footnote{The
same procedure indicates that sources for AdS$_5$ vectors must
satisfy $\Delta>2$, and even higher-spin objects cannot be used at
all. It is not unreasonable to contemplate such deformations, {\it a
priori}, since  sourcing only the $r$ polarizations of fields with
indices in AdS$_5$ would still be consistent with 4D Lorentz
invariance.}  In contrast to the radial wave functions of massive
excitations which are peaked in the IR region, (\ref{eq:scalar_corr}) 
is peaked in the UV.



\section{Evaluating the Angular Integrals}
\label{sec:angints}

The existence of a particular decay channel depends upon having a
nonvanishing overlap of  $T^{1,1}$ harmonics for the various fields
contributing to the effective 4D interaction.  For reference we
give some of the scalar harmonics (we do not know explicit expressions
for vector and tensor harmonics). They have been
calculated in refs.~\cite{bib:harm1,bib:harm2}:
\be
Y_{L}\left(\Psi\right) = J_{l_{1}, m_{1}, R}\left(\theta_{1}\right)
J_{l_{2}, m_{2}, R}\left(\theta_{2}\right) 
e^{\i m_{1}\phi_{1}+\i m_{1}\phi_{1}}
e^{\frac{\i}{2}R\psi}\,,
\ee
The functions $J$ are given by hypergeometric functions,
\baray
J^{\Upsilon}_{l, m, R}\left(\theta\right) &=& N_{L}^{\Upsilon} 
\left(\sin\theta\right)^{m}\left(\cot\frac{\theta}{2}\right)^{\frac{R}{2}}
{_{2}F_{1}}\left(-l+m, 1+l+m, 1+m-\frac{R}{2};
\sin^2\frac{\theta}{2}\right)
\nn\\
J^{\Omega}_{l, m, R}\left(\theta\right) &=& N_{L}^{\Omega} 
\left(\sin\theta\right)^{m}\left(\cot\frac{\theta}{2}\right)^{\frac{R}{2}}
{_{2}F_{1}}\left(-l+\frac{R}{2}, 1+l+\frac{R}{2}, 1-m+\frac{R}{2};
\sin^2\frac{\theta}{2}\right)\nn\\
\earay
where $J^{\Upsilon}$ is nonsingular for $m \ge R/2$ and $J^{\Omega}$ 
is nonsingular for $m \le R/2$. The particular scalar harmonics that 
come into play for the states we are interested in correspond to
$l=1$, $m=0$, $R=0$. The value of $m$ can be inferred following the 
calculation of ref.~\cite{bib:ceresole} where the parameters $r$ and $q$ were
defined in terms of the $m_{1}$ and $m_{2}$ ``magnetic'' quantum
numbers for each of the $SU\left(2\right)$ groups ($r = m_{1} -
m_{2}$, $q = m_{1} + m_{2}$). $q=0$ for scalars, so for the state
$(j,l,r)=(1,0,0)$ we have $m_1=m_2=0$ and the scalar harmonic
\be
Y_{(1,0,0)}\left(y\right) = J_{1, 0, 0}\left(\theta_{1}\right)
J_{0, 0, 0}\left(\theta_{2}\right) = N_{L}^2 \cos\theta_{1}
\ee
 Some  other common harmonics are given by
\bea
Y_{(0,1,0)}\left(y\right) &=& N_{L}^2 \cos\theta_{2}\nn\\
Y_{(1,1,0)}\left(y\right) &=& N_{L}^2 \cos\theta_{1}\cos\theta_{2}\nn\\
Y_{(2,1,0)}\lb y\rb &=& \frac14 N_L^2 \ls\cos\theta_2+\frac32\cos(2\theta_1+\theta_2)+\frac32\cos(2\theta_1+\theta_2)\rs\,.
\label{eq:explharm}
\eea

To evaluate the various integrals we also require the expression of 
$\sqrt{g}$ where $g$ is the metric on the $T^{1,1}$.
If each $SU(2)$ has Euler-angle coordinatization
$(\theta_i,\phi_i,\gamma_i)$, and the left-coset acts to mod out the
$\gamma_i$'s ($\psi=\frac{1}{\sqrt{2}}\lb\gamma_1-\gamma_2\rb$), the
metric in the basis $[\psi,\theta_1,\theta_2,\phi_1,\phi_2]$ is
\be
g_{ab}=\left(\begin{array}{ccccc}
\frac{1}{9} & 0 & 0 & \frac{\cos\theta_{1}}{9} & 
\frac{\cos\theta_{2}}{9}\\
0 & \frac{1}{6} & 0 & 0 & 0 \\
0 & 0 & \frac{1}{6} & 0 & 0 \\
\frac{\cos\theta_{1}}{9} & 0 & 0 & 
\frac{1}{9}\left(1+\frac{\sin^2\theta_{1}}{2}\right) &
 \frac{\cos\theta_{1}\cos\theta_{2}}{9} \\
\frac{\cos\theta_{2}}{9} & 0 & 0 &  
\frac{\cos\theta_{1}\cos\theta_{2}}{9} &
\frac{1}{9}\left(1+\frac{\sin^2\theta_{2}}{2}\right) \\
\end{array}\right)
\ee
with the determinant 
\be
\sqrt{g} =
\frac{\left|\sin\theta_{1}\right|\left|\sin\theta_{2}\right|}
{\sqrt{11664}}\,,
\ee
and coordinate ranges
\bea
\theta_i\in\lb0,2\pi\rb\,,\,\beta_i\in\lb0,\pi\rb\,,\,\gamma_i\in\lb0,4\pi\rb\,.
\eea

%


\end{document}